\documentclass[epj]{svjour}

\usepackage{graphics, color}
\usepackage{bm}       
\usepackage{amsmath}  
\usepackage{amssymb}
\usepackage{mathtools}
\usepackage{braket}
\usepackage{footnote}
\begin{document}
\title{Structure models: from shell model to ab initio methods}
\subtitle{A brief introduction to microscopic theories for exotic nuclei}
\author{Sonia Bacca\inst{1,2}
}                     
%
%
\institute{TRIUMF, Vancouver,  British Columbia, V6T 2A3, Canada \and
Department of Physics and Astronomy, University of Manitoba, Winnipeg, MB, R3T 2N2, Canada}
\date{Received: date / Revised version: date}
%
\abstract{A brief review of models to describe nuclear structure and reactions properties is presented, starting from the historical shell model picture and encompassing modern ab initio approaches. A selection of recent theoretical results on observables for exotic light and medium-mass nuclei is shown. Emphasis is given to the comparison with experiment and to what can be learned about three-body forces and continuum properties. 
} 
\maketitle
\section{Introduction}
\label{intro}

An atomic nucleus of mass number $A$ consists of a set of $Z$ protons and $N$ neutrons, strongly interacting with each other. This is the basic picture of the nucleus, which will be discussed in this work. While the strong and electroweak forces govern nuclear properties, the fundamental theory of quantum-chromo-dynamics  does not admit a simple solution at the low-energy scales of few to several MeV relevant to nuclear physics. Thus, nuclear interactions  are typically modeled in terms of effective forces among the relevant  degrees of freedom, {\it i.e.}, the nucleons.

The understanding of structure and reaction properties of nuclei has been the center of theoretical and experimental studies since the
beginning of nuclear physics, about 100 years ago. Models and theories were built to describe experimental observations, starting from simple and intuitive ones and going on to more sophisticated methods. Today,  the goal of modern nuclear theory is to describe nuclear properties in terms of  theories which are rooted as much as possible in the fundamentals of quantum-chromo-dynamics. 

While the direct solution of the nucleus starting from fundamental degrees of freedom, quarks and gluons, is being pursued for light nuclei, see, {\it e.g.}, Refs.~\cite{Bean2014,Bean2013}, this description is still in its infancy stage and systematic errors are quite large so that a comparison with experimental data is sometimes difficult. On the other hand, enormous progress has been made in describing light and medium-mass nuclei with inter-nucleon forces that are inspired by the symmetries of quantum-chromo-dynamics and  well-constrained by experimental data~\cite{Epelbaum06,Machleidt11,Epelbaum12}. Several so called ab initio methods can describe nuclei with increasing mass number, see, {\it e.g.}, Refs.~\cite{Leidemann2013,Bacca2014,Hebeler2015} for recent reviews, with an accuracy sometimes comparable to that of experimental data. An estimate of the systematic uncertainty in light nuclei suggests that this theoretical picture has the right accuracy to solve longstanding problems in nuclear physics~\cite{Binder2015}.  Hence, it is  at the moment  our best chance  to provide guidance and help interpret experiments performed at the rare isotope facilities, where exotic nuclei far from the stability line are studied.

In this work, we will present some of the theoretical models for structure and reactions developed for nuclei. We will  review the simple shell model picture, then discuss more complex ab initio descriptions of atomic nuclei.

\section{The shell model in nuclei}

\begin{figure}
\centering
  \includegraphics[width=\linewidth]{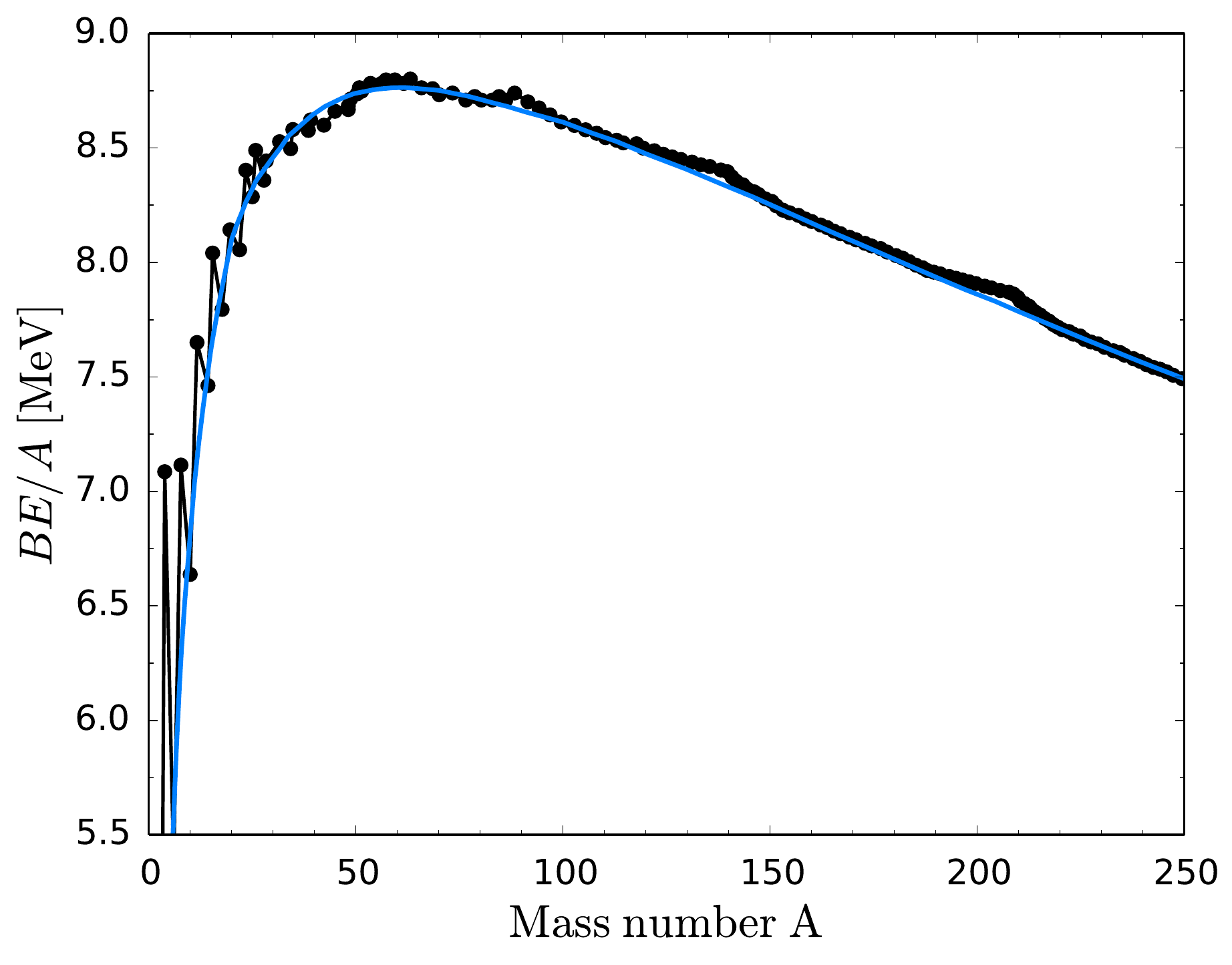}
  \caption{Binding energy per nucleon as a function of mass number $A$. Dots correspond to experimental data  for stable nuclei and the curve shows the behavior of the empirical formula of Eq.~(\ref{semi-emp}).}
\label{be_fig}
\end{figure}

The first observable that one would like to describe for nuclei is their binding energy. The latter is the energy difference between the constituents -- protons and neutrons -- and the compound object -- the nucleus --, defined as
\begin{equation}
BE(Z,A)=Zm_pc^2+Nm_nc^2-m_N(Z,N)c^2 \,,
\label{be}
\end{equation}
where $m_{p,n}$ is the mass of the proton or neutron and $m_N(Z,N)$ is the mass of the nucleus.
Data on the behavior of the binding energy per nucleon as a function of mass number are shown in Figure~\ref{be_fig}.
One clearly observes that the curve is almost constant for $100 \lesssim A \lesssim 250$, located at an average value of about 8 MeV, and while it is mostly smooth, especially at large mass number, it also presents peaks and structures.

Given that $BE(Z,A)$ is the simplest observable one can study, several theoretical models to describe it were  developed.
The simplest model, based on an analogy between nuclei and liquids  -- thus also called droplet model -- was proposed by von Weisz\"{a}cker~\cite{weisz}. The model lead to a formula where the binding energy is described as a function of $Z$ and $A$ by
\begin{equation}
\label{semi-emp}
BE(Z,A)=a_{vol} A - a_{sur}A^{2/3} -a_{Coul}\frac{Z^2}{A^{1/3}} -a_{asy} \frac{(N-Z)^2}{A}\,.
\end{equation}
The  four terms are named volume, surface, Coulomb and asymmetry term, respectively, and the coefficients $a_{vol}$, $a_{sur}$, $a_{Coul}$ and $a_{asy}$ are typically  fit to experimental data of stable nuclei. The analogous formula which is obtained for nuclear masses by combining Eq.~(\ref{be}) and (\ref{semi-emp}) is called semi-empirical mass formula.
While such model describes quite well the binding energy per particle over the mass range, it completely fails to reproduce the peaked structures in  Figure~\ref{be_fig}.

Non smooth and discontinuous behaviors in observables may be an indication of the presence of shell structure.
For example, in atomic systems, it is observed that energies or radii as a function of proton number $Z$ present discontinuities
due to the filling of atomic shells, see, {\it e.g.}, Ref.~\cite{Krane}. 
The shell model for atoms has proven to be a very simple theory that can account for observed properties.
Consequently, it is natural to ask the question of whether nuclei also exhibit  shell structure.
A positive answer is in principle not so obvious. In fact, while the atom has a natural center -- the nucleus -- where all electrons
orbit around, the nucleons do not have an analogous center they orbit around. Moreover, while in case of atoms the nucleus provides a Coulomb attractive
external potential which holds the system together, there is  no external potential within the nucleus. Thus, one might be tempted to think that there should not be any shell structure in nuclei. 

\begin{figure}
\centering
  \includegraphics[width=\linewidth]{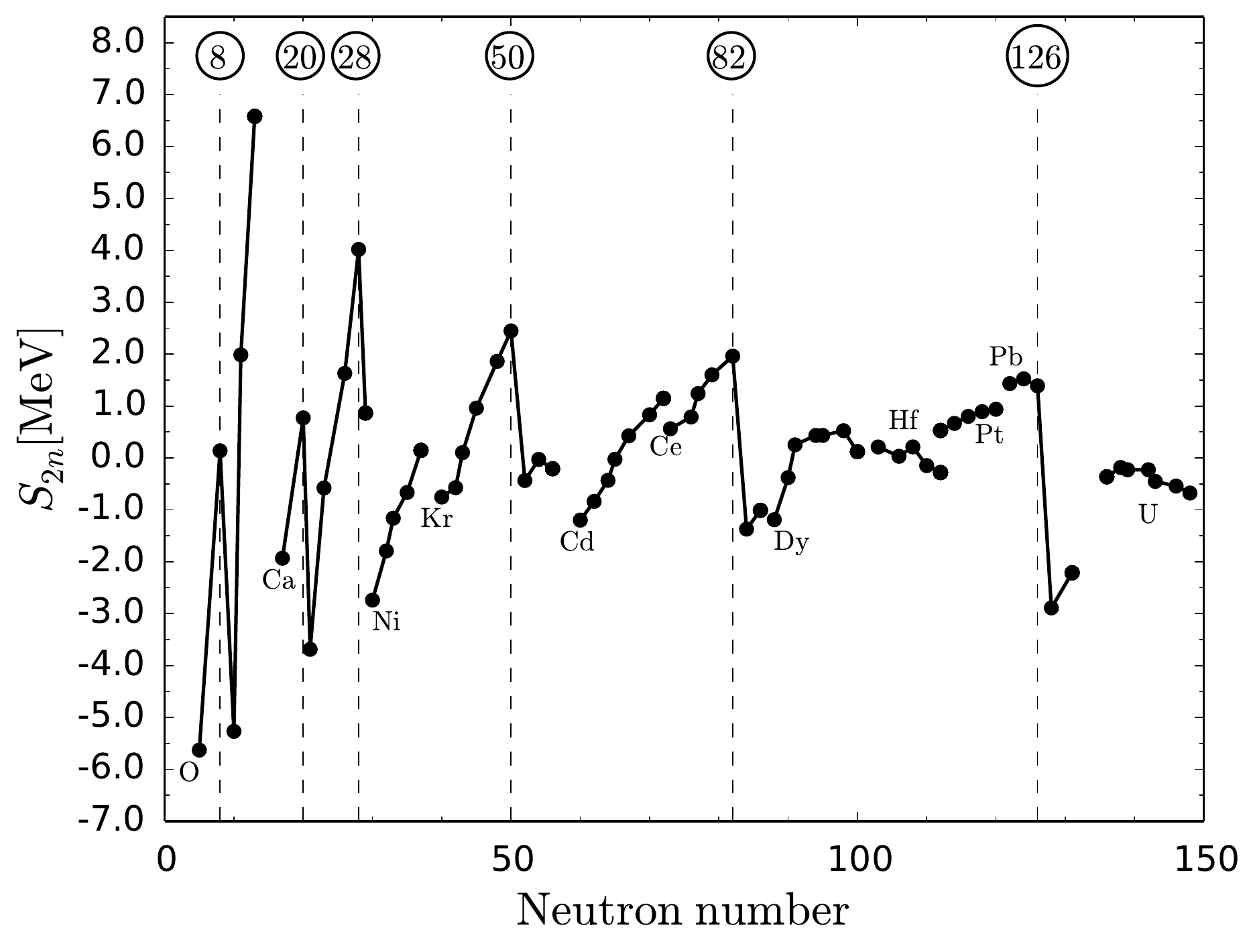}
  \caption{Two-neutron separation energies for different isotopic chains as a function of neutron number $N$: difference between experimental data and the prediction of the empirical mass formula. Magic numbers are indicated with circles in correspondence of the sudden changes of this observable. Data taken from Ref.~\cite{Krane}.}
\label{s2n_fig}
\end{figure}
\begin{figure}
\centering
  \includegraphics[width=\linewidth]{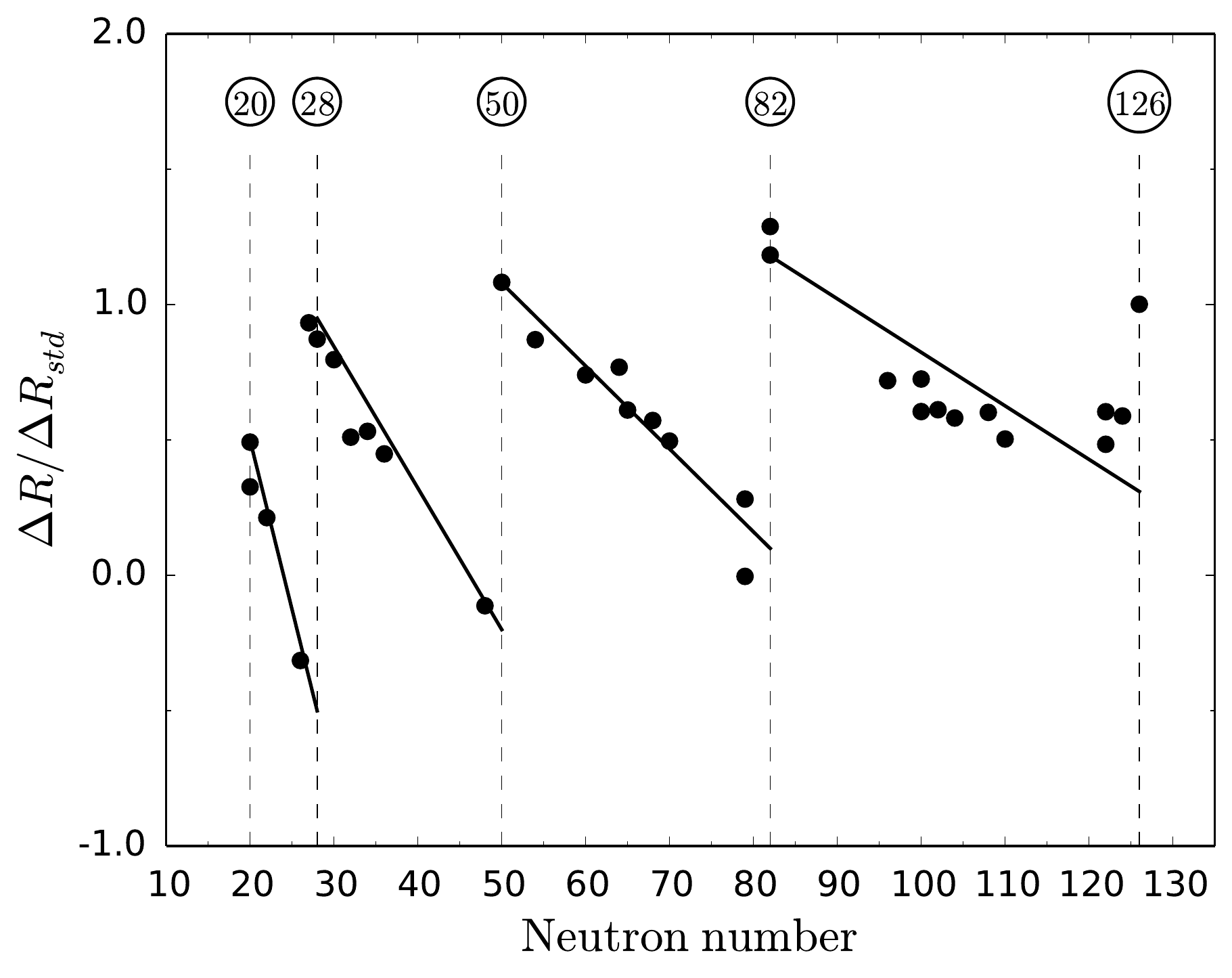}
  \caption{Nuclear charge radii as a function of neutron number $N$: difference between experimental data and the prediction of the droplet model. Magic numbers are indicated with circles in correspondence of the sudden changes of this observable. Data taken from Ref.~\cite{Krane}.}
\label{r_fig}
\end{figure} 

However, experimental evidence says otherwise.
In Figures~\ref{s2n_fig} and \ref{r_fig},  separation energies and radii are shown, respectively,  for different nuclei as a deviation from the behavior described by the droplet model.  The separation energy, similarly to the ionization energy in atoms,
 is defined as a difference of binding energies. In particular, the two-neutron separation energy is
\begin{equation}
S_{2n}=BE(Z,A)-BE(Z, A-2)\,,
\end{equation} 
where from the starting nucleus with $Z$ protons and $N$ neutrons, one subtracts the binding energy of a nucleus with two less neutrons, {\it i.e.}, $Z$ protons and $N-2$ neutrons, thus $A-2$ nucleons in total.
Figure~\ref{s2n_fig}  shows that $S_{2n}$ increases with  $N$, a part from sudden drops that occur at specific values of $N$.  

Figure~\ref{r_fig} shows the change of the nuclear charge radius with respect to the $A^{1/3}$ dependence expected from the droplet model. Also in this case a discontinuous behavior in proximity of the very same values of neutron numbers is seen.
Observations made regarding both Figure~\ref{s2n_fig} and \ref{r_fig}, together with the non-smooth behavior of the $BE(Z,A)$ function in Figure~\ref{be_fig}, indicate indeed the presence of shell structures in nuclei. The values of neutron number $N$ (and similarly of proton number $Z$) where one observes these discontinuous patterns  -- the equivalent of atomic shell closures -- are called ``magic numbers'' in nuclear physics.

In order to explain what magic numbers are, one has to abandon the simple liquid drop model and try to construct a more microscopic theory, where the nucleus is described in terms of a collection of protons and neutrons.

Let us start building a microscopic theory for the nucleus by considering that,
at very low-energy, it is fair to treat the nucleons as point-like particles and neglect their internal structure in terms of quarks and gluons. Moreover,
given the lengths scale of the atomic nucleus, a quantum mechanical treatment should be used.
In a non-relativistic framework, justified by the fact that nucleons are almost 1~GeV heavy and do not move quite at the speed of light,  a microscopic description of the nucleus requires to solve the Schr\"{o}dinger equation 
\begin{equation}
H\Psi = E \Psi \,.
\label{schro}
\end{equation}
Here, the Hamiltonian 
\begin{equation}
H=K+V=\sum_i^A \frac{p_i^2}{2m} +V
\label{hamilt}
\end{equation}
 includes the kinetic energy $K$, with $m$ being the mass of the nucleon (assuming $m_p=m_n$), and the potential $V$. Its eigenstates
$\Psi=\Psi ({\bf r}_1, {\bf r}_2, \dots, {\bf r}_A)$ are many-body wave functions, with  ${\bf r}_i$ being the coordinate of the $i$-th particle in the laboratory frame. For simplicity, we are omitting spin and isospin degrees of freedom.
The potential $V$ will be describing the strong interaction among nucleons, and for protons it would obviously include the Coulomb force as well.
What determines the complexity and sophistication of this theoretical description is the model used for the interaction $V$ and the way
the many-body wave function  $\Psi$ is constructed. 

\section{Non-interacting shell model}

The simple non-interacting shell model is based on a mean-field {\it ansatz}, namely each nucleon is assumed to be moving in an external field created by the remaining $A-1$ nucleons~\cite{Suhonen}. Such a mean-field potential can be interpreted as a time-average of the interactions of each nucleon with the neighbors and can be written as a one-body potential
\begin{equation}
V_{MF}=\sum_i^A v({\bf r}_i)\,.
\end{equation} 
Under this approximation the problem of $A$ strongly interacting nucleons becomes a problem of $A$ non-interacting particles
under the influence of an external field $v$.  The corresponding  many-body Schr\"{o}dinger equation  is 
\begin{equation}
\left[\sum_i^A  \frac{p_i^2}{2m}+  v({\bf r}_i)\right] \Psi  = E \Psi  \,,
\label{smf}
\end{equation}
whose solution can be found by assuming that the many-body state is a product of single-particle states 
\begin{equation}
\Psi ({\bf r}_1, {\bf r}_2, \dots, {\bf r}_A) = \phi ({\bf r}_1) \phi({\bf r}_2) \dots  \phi({\bf r}_A)=\prod_{i=1}^A  \phi({\bf r}_i)\,,
\label{prod}
\end{equation}
given that the Hamiltonian does not contain any interaction between particles.

Using Eq.~(\ref{prod}) and substituting it in Eq.~(\ref{smf}), one obtains the following single-particle  Schr\"{o}dinger equation
\begin{equation}
h({\bf r})  \phi_k({\bf r}) = \left[ \frac{p^2}{2m} +  v({\bf r})\right] \phi_k ({\bf r})= \varepsilon_k   \phi_k({\bf r})\,,
\label{sp}
\end{equation}
where the index $k$ labels eigenstates and eigenfunctions of the one-body problem, represented here by the coordinate ${\bf r}$ (and conjugate momentum ${\bf p}$) of one particle. 

\begin{figure}
\centering
  \includegraphics[width=0.7\linewidth]{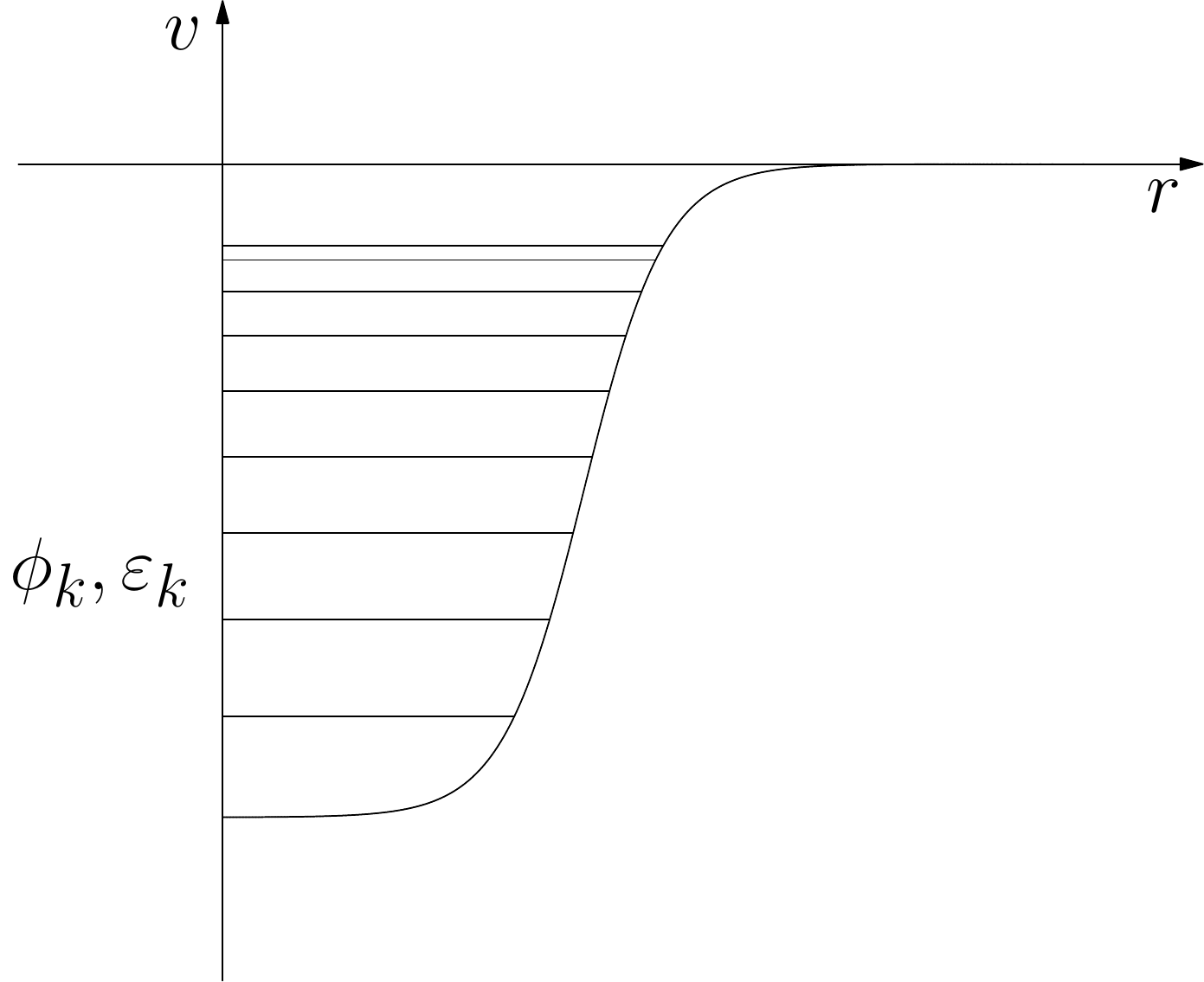}
  \caption{Schematic representation of a mean-field potential with its single-particle states $\phi_k$ and energies $\varepsilon_k$.}
\label{ws_fig}
\end{figure}

If the mean-field potential $v$ is known, the solution
of Eq.~(\ref{sp}) is quickly found. Note that solutions could be either analytical or numerical, depending on the choice of the mean-field potential $v$. A schematic representation of the mean-field potential and its single-particle states $\phi_k$ -- also called orbitals -- with corresponding energy levels $\varepsilon_k$ is shown in Figure~\ref{ws_fig}.
Finally, the solution of the many-body problem in Eq.~(\ref{smf}) is
\begin{equation}
E=\sum_{k=1}^A \varepsilon_{k} d_{k}\,,
\end{equation}
where $d_{k}$ is the degeneracy of the energy level, which can be different from one.

In essence, in a mean-field approximation, the solution of the many-body problem is simply found by solving the single-particle  Schr\"{o}dinger equation
with an external potential $v$.

Since nucleons are fermions and the Pauli principle should be respected, actually Eq.~(\ref{prod}) should be modified to include an antisymmetrizer operator ${\mathcal A}$, which performs  permutations of particles accompanied by a sign.
Thus, the appropriate {\it ansatz} for the many-body wave function is
\begin{equation}
\Psi ({\bf r}_1, {\bf r}_2, \dots, {\bf r}_A) = {\mathcal A} \left[ \prod_{i=1}^A  \phi({\bf r}_i) \right]\,.
\label{prodA}
\end{equation}
 
The antisymmetrized product state is called  Slater determinant. In fact, it can be easily calculated as the determinant of a matrix 
\begin{equation}
\Psi ({\bf r}_1, {\bf r}_2, \dots, {\bf r}_A)=\frac{1}{\sqrt{A!}} {\rm det}\!\! \left(
\begin{array}{cccc}
\phi_1 ({\bf r}_1)& \phi_1 ({\bf r}_2)&\dots& \phi_1 ({\bf r}_A)\\
\phi_2 ({\bf r}_1)& \phi_2 ({\bf r}_2)&\dots& \phi_2 ({\bf r}_A)\\
\vdots & \vdots & \ddots &\vdots\\
\phi_A ({\bf r}_1)& \phi_A ({\bf r}_2)&\dots& \phi_A ({\bf r}_A)\\
\end{array}
 \right)\!\!,
\end{equation}
where every row contains the same single-particle state and every column refers to the same particle.
A Slater determinant is the many-body solution of a mean-field potential and is a very general concept, also used in modern methods.

The shell model appeared first in the 1920s, but at the beginning it was unsuccessful in explaining observations, in particular in
describing the location of the magic numbers. No matter what mean-field potential $v$ was chosen, {\it e.g.}, harmonic oscillator, square-well or Wood-Saxon potential, predicted magic numbers would not correspond to observation, see, {\it e.g.}, Ref.~\cite{Krane}. 

\begin{figure}
\centering
  \includegraphics[width=0.6\linewidth]{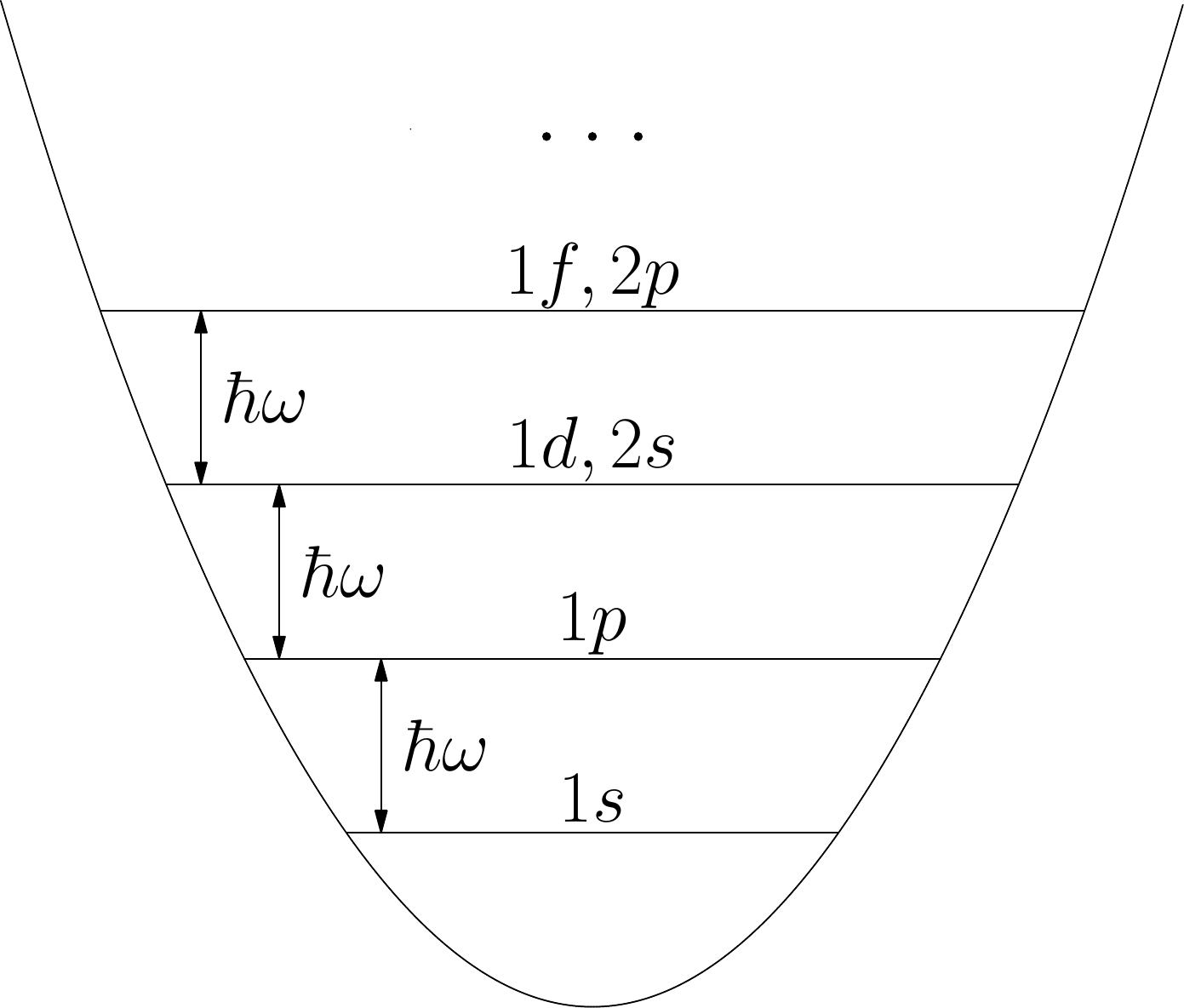}
  \caption{Schematic representation of the harmonic oscillator mean-field model.
  The spectroscopic notation is used to denote single-particle states, where the letter indicates the orbital angular momentum ($s$ corresponds to $\ell=0$, $p$ to $\ell=1$, $d$ to $\ell=2$, etc.).}
\label{ho_fig}
\end{figure}

Let us consider the simple example of a harmonic oscillator potential. In this case Eq.~(\ref{sp}) becomes
\begin{equation}
\left[\frac{p^2}{2m} + \frac{1}{2} m \omega^2 r^2 \right]  \phi_k({\bf r}) = \varepsilon_k   \phi_k({\bf r})\,,
\label{sp_ho}
\end{equation}
where we know the analytical solution to be
\begin{equation}
 \varepsilon_k =  \varepsilon_{\mathcal N}= \left({\mathcal N}+\frac{3}{2}\right) \hbar\omega \,.
\end{equation}
Here ${\mathcal N}$ is equal to
$2n+\ell$, where $n$ is the nodal quantum number,  $\ell$ is the orbital quantum number and the degeneracy factor is
\begin{equation}
 d_k =  d_{\mathcal N}= 2(2\ell+1)\,.
\end{equation}
In the last  equation  the factor 2 comes from the possibility of having a nucleon with spin up or down, while the $(2\ell+1)$--factor arises from the $m_\ell$ degeneracy,  where $m_\ell$ is the projection of the orbital angular momentum $\ell$. 
A schematic representation of the energy levels for a harmonic oscillator mean-field potential is shown in Figure~\ref{ho_fig}.

\begin{center}
\begin{table}[bth]
\centering{
\caption{Quantum numbers, energy and expected magic numbers ($\sum_{\mathcal N}d_{\mathcal N}$) obtained from a harmonic oscillator mean-field potential. The spectroscopic notation is used to denote single-particle states (last column).}}
\begin{tabular}{c|c|c|c|c}
\label{ho_tab1}
${\mathcal N}$& $E_{\mathcal N}$& $d_{\mathcal N}$& $\sum_{\mathcal N} d_{\mathcal N}$& orbital\\
\hline
0 & $3/2~\hbar \omega$& 2& 2& 1$s$\\
1 & $5/2~\hbar \omega$& 6& 8& 1$p$\\
2 & $7/2~\hbar \omega$& 12& 20& 1$d$,~2$s$\\
3 & $9/2~\hbar \omega$& 20& 40& 1$f$,~2$p$\\
\dots & $\dots$& $\dots$& $\dots$& $\dots$\\
\end{tabular}
\end{table}
\end{center}

If we want to describe the ground-state of a nucleus, we start from Figure~\ref{ho_fig} and  fill in the orbitals with as many nucleons as allowed by the degeneracy factor. Table 1 
 shows what the magic numbers are in this model. They are given by the total numbers of particles that correspond to completely full orbitals, {\it i.e.} $\sum_{\mathcal N} d_{\mathcal N}$.   Because in nuclear physics one has two kinds of nucleons -- protons and neutrons -- with a different mean field mostly due to the Coulomb force acting only between  protons, we will have two kinds of magic numbers: those for protons and those for neutrons. Looking at Table 1, the  magic numbers predicted by this model are $Z$ or $N=$ 2,8, 20, 40, $\dots$.
The nucleus with $N$ or $Z$ corresponding to a full shell will have a large energy gap -- in this case $\hbar\omega$ -- with respect to its mass neighbors, and as such it will be a ``magic'' nucleus.
From a close look at Figures~\ref{s2n_fig} and \ref{r_fig}, one sees that only the first three magic numbers correspond to observations, but not the subsequent ones.

\begin{figure}
\centering
  \includegraphics[width=0.9\linewidth]{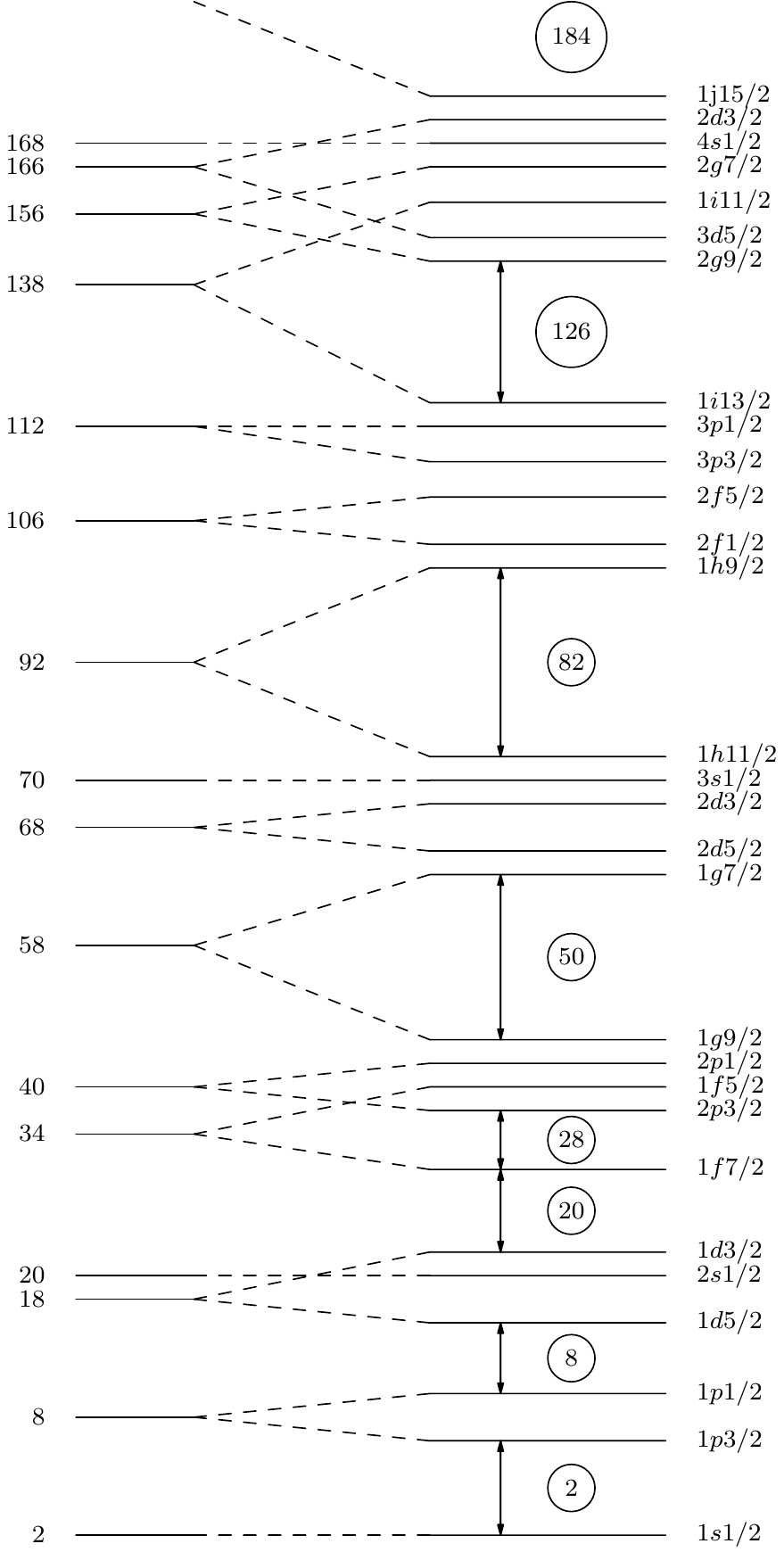}
  \caption{Energy levels and shell structure with magic numbers (highlighted with circles) obtained with a mean-field model that contains a spin-orbit force. The degeneracy factor of this model is $(2j+1)$. The spectroscopic notation is used on the right to describe the single-particle state or orbital, with the first number on the left counting the occurrences of each state and the number on the right indicating the total angular momentum $j$ associated to that state.}
\label{sm_level}
\end{figure} 

  Only in 1945, with the introduction of a spin-orbit term in the mean-field potential by Maria Goppert Mayer \cite{GoppertMayer} and Hans Jensen~\cite{MayerJensen}, it was established that
the shell model was an important tool to describe nuclear physics. Goppert Mayer and Jensen were awarded the Nobel prize in 1963.

In the presence of a spin-orbit component in the mean-field potential, such as
\begin{equation}
v_{\ell\cdot s}(r)~ {\boldsymbol \ell} \cdot {\bf s}\,,
\end{equation}
it is appropriate to introduce the total angular momentum as a conserved quantum number ${\bf j}={\boldsymbol \ell}+ {\bf s}$.
The spin-orbit operator can be then written as
\begin{equation}
 {\boldsymbol \ell} \cdot {\bf s}=\frac{1}{2}(j^2-\ell^2 -s^2)\,,
\end{equation}
and the expectation value  of this operator on single-particle states will be
\begin{equation}
 \langle {\boldsymbol \ell} \cdot {\bf s} \rangle=\frac{1}{2}(j(j+1)-\ell(\ell+1)- 3/4)~\hbar^2\,.
\end{equation}
The effect of the spin-orbit term is that every single-particle level with angular momentum different from zero gets split into two states:
one with total angular momentum $j=\ell-1/2$ and one with $j=\ell+1/2$. 
In nuclear physics, experimental evidence shows that the spin-orbit force is attractive, thus $v_{\ell\cdot s}$ is taken to be negative and the spin-orbit splitting is such that the state with  $j=\ell+1/2$ is found at a lower energy with respect to the  state with  $j=\ell-1/2$.
 When one calculates the energy difference between these two states, one obtains
\begin{equation}
 \langle {\boldsymbol \ell} \cdot {\bf s} \rangle_{j=\ell+1/2} -\langle {\boldsymbol \ell} \cdot {\bf s} \rangle_{j=\ell-1/2}=  \frac{1}{2}(2\ell+1)~\hbar^2\,.
\end{equation}
One observes that the above energy splitting increases with increasing orbital angular momentum $\ell$. This is  evident in Figure~\ref{sm_level}, where the shell structure of a mean-field model with spin-orbit force is shown.

Magic numbers, obtained by filling each orbital with all the nucleons allowed by the degeneracy factor $(2j+1)$, are the numbers of $N$ and $Z$ for which the last orbital is full and the next one is found at higher energy with a large energy gap. In Figure~\ref{sm_level},  energy gaps are indicated by arrows and magic numbers are highlighted with a circle.
They are predicted to be $N$ or $Z=2, 8,20, 28,50$ and $82$, which correspond to observation. Experimental evidence shows that in case of the neutrons there is an additional magic number $N=126$ which is not observed for protons as we do not have elements with that high $Z$, due to the  strong Coulomb force, that prevents nuclei to be bound. This magic number is also predicted by the model.

Single-particle states  in Figure~\ref{sm_level} are labeled on the right with the spectroscopic notation. The latter is similar to what explained in Figure~\ref{ho_fig}, but it also has the $j$ quantum number indicated  on the right.
Typically each single-particle state constitutes a sub--shell, while a full shell is determined by a group of orbitals  separated from the others by  a large energy gap. A clarifying example is given by the $1p_{3/2}$ and $1p_{1/2}$ sub-shells, originated by the splitting due to the spin-orbit force, which together form the whole $p$--shell.

\begin{figure}
\centering
  \includegraphics[width=0.8\linewidth]{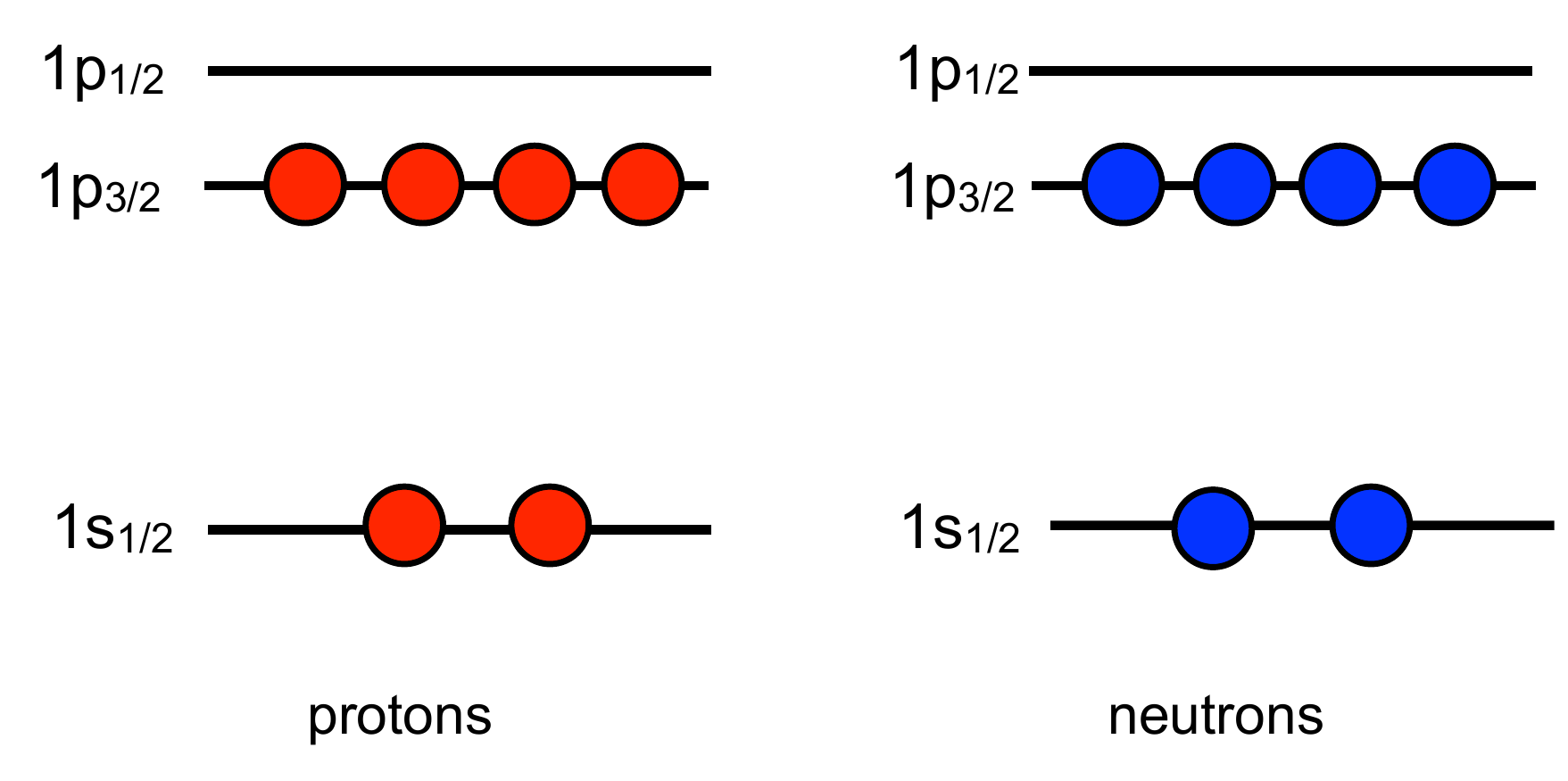}
  \caption{Non-interacting shell model configuration for the nucleus of $^{12}$C, where the $1s_{1/2}$ and the $1p_{3/2}$ levels are fully occupied by both the $6$ protons and $6$ neutrons.}
\label{C12_fig}
\end{figure}

Using single-particle energy levels shown in Figure~\ref{sm_level} one can  easily construct the ground-state as well as excited-states for a variety of nuclei. In Figure~\ref{C12_fig} we show, as an example,  the ground-state of the $^{12}$C nucleus, where the $6$ protons and $6$ neutrons fill  the $1s_{1/2}$ and the $1p_{3/2}$ energy levels. Excited-states of nuclei can be constructed by particle-hole excitations, {\it i.e.} by 
removing  nucleons from the lowest levels and promoting them to  higher states.

\section{Interacting shell model}
\label{ism}
The non-interacting shell model is a  simple theory which  accounts for some observations, but is still a crude approximation of the full problem  of describing the nucleus formulated in  Eq.~(\ref{schro}). This is  mostly due to the fact that  particles are assumed  not to interact with each other.
In reality, the potential $V$ of Eq.~(\ref{hamilt}) is not a simple mean-field potential $V_{MF}$, but it contains pairwise interactions, named two-body forces, and in general also many-body terms, such as
\begin{equation}
V=\sum_{i<j}^AV_{ij}+\sum_{i<j<k}^AW_{ijk}+ \dots \,.
\label{pot}
\end{equation} 

\begin{figure}
\centering
  \includegraphics[width=0.5\linewidth]{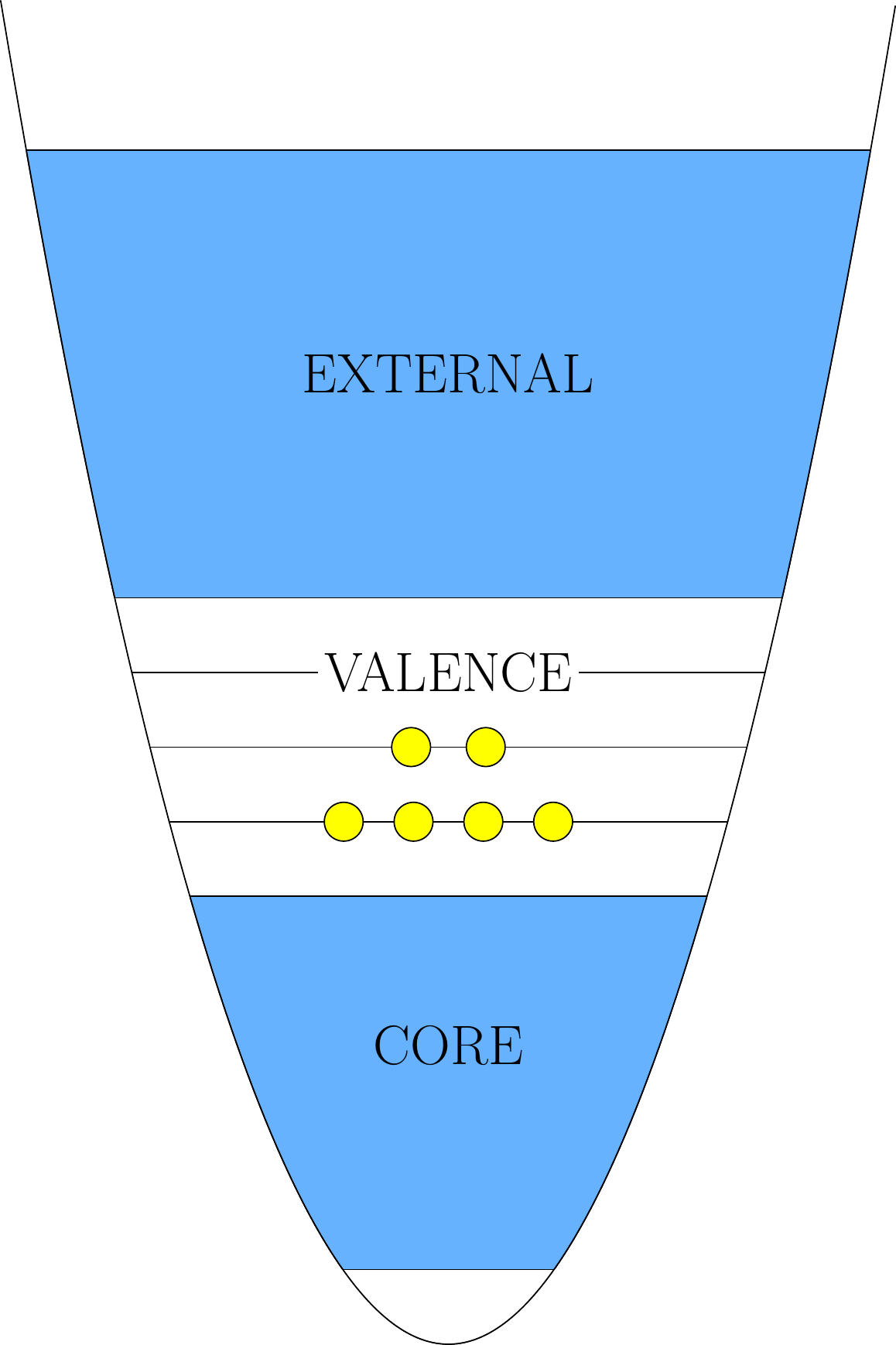}
  \caption{Visualization of the interacting shell-model picture, where orbitals are separated into core, valence space and external space.}
\label{gabriel}
\end{figure}

Let us first assume that there are only two-body forces $V_{ij}$, which depend on coordinate ${\bf r}_i$ and ${\bf r}_j$ and consider the following Hamiltonian
\begin{equation}
H=\sum_i^A \frac{p_i^2}{2m} + \sum_{i<j}^AV({\bf r}_i, {\bf r}_j)\,.
\label{hamilt2}
\end{equation}
Summing and subtracting a mean field potential as
\begin{eqnarray}
\nonumber
H&=& \left [\sum_i^A \frac{p_i^2}{2m} + \sum_i^A v({\bf r}_i) \right]+\\
\nonumber
&+& \left [ \sum_{i<j}^AV({\bf r}_i, {\bf r}_j)  - \sum_i^A v({\bf r}_i) \right]=\\
&=&H^0+W_{\rm RES}\,,
\label{h_sep}
\end{eqnarray}
one can separate the Hamiltonian into a non-interacting part $H^0$ and a residual interaction $W_{\rm RES}$.
If the residual interaction is small, the problem is well-approximated by a mean-field solution or by perturbations around it.
When the residual interaction is not small, corrections to the mean-field solutions can be obtained non-perturbatively by diagonalizing the whole Hamiltonian $H$ on a basis of eigenstates of $H^{0}$.
This is the main idea behind the interacting shell model.

In order to start an interacting shell model calculation, one first has to make an {\it ansatz} for the form of the mean-field potential.
Let us consider the harmonic oscillator potential, for simplicity, but any mean-field choice is valid. Secondly, all the single-particle states are assumed to be separated into:
\begin{itemize}
\item inert core;
\item valence space;
\item external space.
\end{itemize}
A pictorial representation  of this separation is shown in Figure~\ref{gabriel}.

\begin{figure}
\centering
  \includegraphics[width=0.8\linewidth]{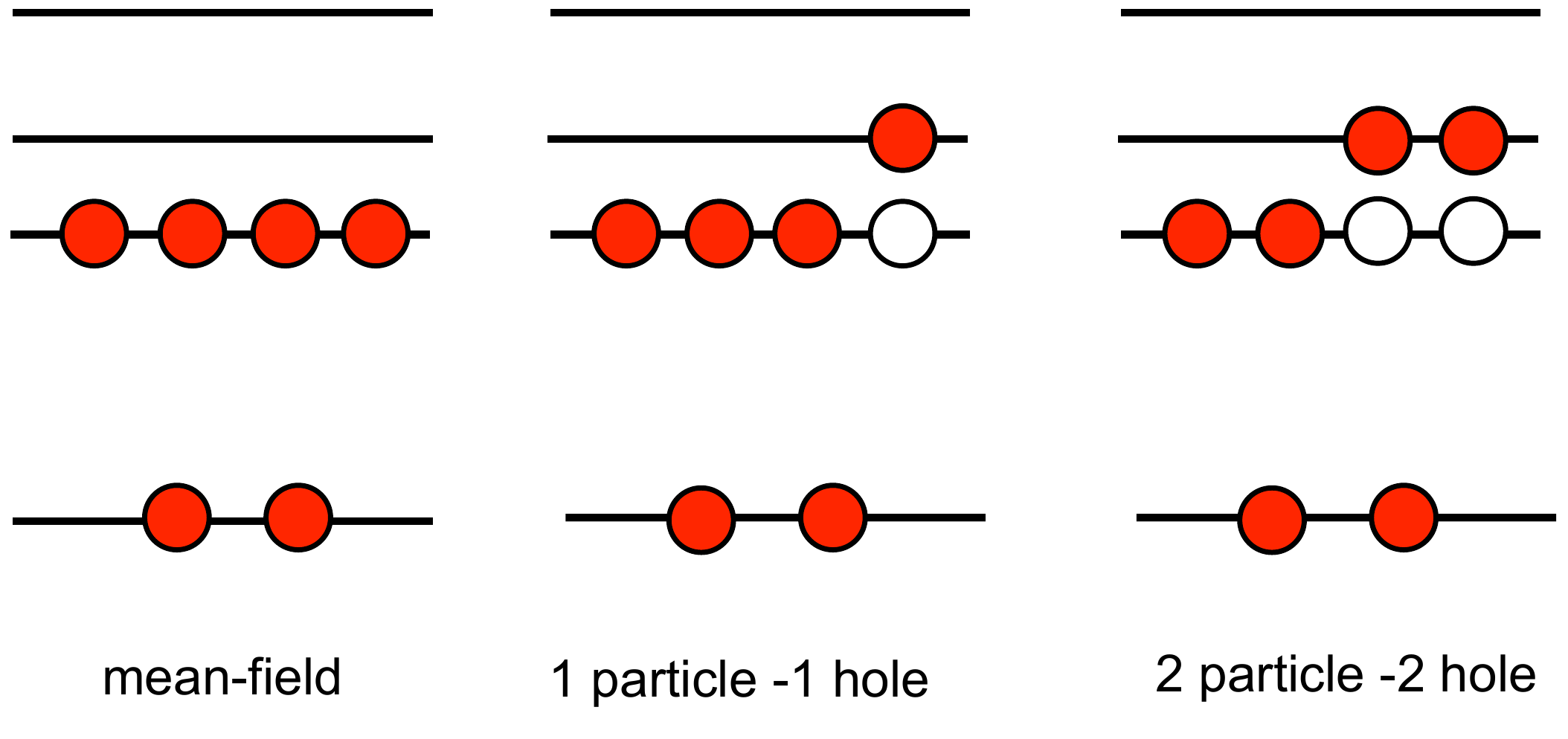}
  \caption{Schematic picture of states generated by particle-hole excitations on top of the mean-field ground-state.}
\label{parthole}
\end{figure}

 The inert core is constituted by orbitals that are always full, while the valence space contains orbitals where one can have particle-hole excitations, as represented in Figure~\ref{parthole}.  Finally, the external space is a collection of orbitals that are always empty.

The next step is to construct single-particle energies $\varepsilon_k$  and effective two-body interactions $V^{\rm eff}_{ij}$ such that the information included in the external space
is  projected into the valence space. This is typically done either by using many-body perturbation theory or by phenomenologically fitting matrix elements to experimental data.  Alternative more modern tools can also be used, see, {\it e.g.}, Ref.~\cite{tsuki,scott,janseng,ragnar}. Then, one diagonalizes the Hamiltonian in Eq.~(\ref{hamilt2}) where $V_{ij}$ is substituted by  $V^{\rm eff}_{ij}$ in the valence space. Basically, one expands $H$ on a set of basis states as
\begin{equation}
\Psi=\sum_{\beta} c_{\beta} \Psi_{\beta} \,,
\label{exp}
\end{equation}
where the states $\Psi_{\beta}$ are obtained by taking particle-hole excitations on top of the mean-field spanning all orbitals in the valence spaces. Each of these states, represented schematically in Figure~\ref{parthole}, is a different Slater determinant. The coefficients $c_{\beta}$ will be the result of the diagonalization.

Several phenomenological interactions exist in the literature which have been constructed for a specific choice of  core and valence spaces. For example, to study $p$-shell nuclei, the Cohen-Kurath interaction~\cite{cohen-kurath} can be used, where the core is made by the $1s_{1/2}$ shell and the valence space is the $p$-shell ($1p_{3/2}$ and $1p_{1/2}$ orbitals). For $sd$-shell nuclei in the mass range $16 \le A \le 40$, the USD interaction~\cite{USD} is widely used. In this case the core is $^{16}$O and the valence space is the $sd$-shell ($1d_{5/2}$, $2s_{1/2}$  and $1d_{3/2}$ orbitals). Finally, for the $pf$-shell nuclei, the GXPF1~\cite{GXPF1} or KB3G~\cite{KB3G}  interactions are commonly used, where the assumed core is $^{40}$Ca and the valence space is the $pf$-shell ($1f_{7/2}$, $2p_{3/2}$, $1f_{5/2}$  and $2p_{1/2}$ orbitals).

The interacting shell model is a very successful theory and is used to understand and interpret experimental data. The interested reader can consult Refs.~\cite{Brown,Poves,BrownRev,Caurier,Otsuka,Coraggio1,Coraggio2} for more detailed information and applications. 

\section{Chiral effective field  theory}
\label{chiral}
While the interacting shell model is frequently used to study structure and reaction properties of nuclei, the phenomenological approach to derive effective forces as a set of matrix elements by fitting to a sample of data, may be lacking the desired  link to quantum-chromo-dynamics. Thus, alternative strategies have been identified to find a deeper connection to the fundamental theory.

The recent history of nuclear physics has witnessed
a tremendous development of effective field theories (EFT)  that 
systematically describe the interactions of nucleons among themselves and
with external probes using effective degrees of freedom. In particular, the chiral EFT ($\chi$EFT), inspired by the 
explicit and spontaneous symmetry breaking of quantum-chromo-dynamics,  allows the construction of 
 interactions and currents, which
preserve all the relevant symmetries, including chiral symmetry.
Potentials and currents are expanded
in powers $\nu$ of $Q/\Lambda_\chi$, where $Q$ is the momentum associated with the observable and $\Lambda_\chi\sim1$ GeV
represents the chiral-symmetry breaking scale. At energies and momenta well below 1 GeV,
 $\chi$EFT provides an expansion  in powers of a small parameter~\cite{Epelbaum12,Machleidt11}  and thus it is expected to converge. The interested reader may consult, {\it e.g.}, Ref.~\cite{Binder2015} for a recent update on the status of chiral convergence.

\begin{figure}
\centering
  \includegraphics[width=0.5\linewidth]{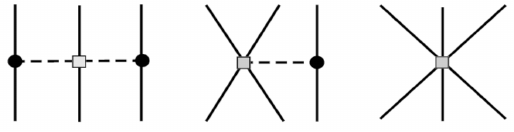}
  \caption{Feynman diagrams of three-nucleon forces at N2LO: solid lines are nucleons and dashed lines are pions. The last two diagrams contain contact terms, a two- and a three-nucleon contact term, respectively, whose low-energy constants are typically calibrated on $A\ge2$ observables.}
\label{3nf}
\end{figure}

The coefficients of the chiral expansion, that appear in this scheme, are called
low-energy constants. They  encapsulate high-energy physics which cannot
be resolved by a low-energy theory. They are unknown and need to be fixed by comparison
with the experimental data. The most common strategy is to first calibrate the nucleon-nucleon (NN) forces at next-to-next-to-next-to-leading order (N3LO) on nucleon-nucleon scattering data and then  tune three-nucleon (3N) forces at next-to-next-to-leading-order (N2LO) on  $A\ge2$ observables, see, {\it e.g.}, Ref.~\cite{Gazit09,Marcucci12}.  Figure~\ref{3nf} shows the Feynman diagrams of 3N forces at N2LO, which are mostly used in the applications to light- and medium-mass nuclei.
Other paradigms are also being explored, where, in particular, a simultaneous optimization of the low energy constants order by order is pursued, rather than by groups of particles, see, {\it e.g.},  Ref.~\cite{andreas}.

Within the same framework of $\chi$EFT it is also possible to derive current operators which describe how the nucleons couple with an external electroweak field. In analogy to potentials, currents are expanded in many-body operators and corrections to the leading order components have been derived for example in Refs.~\cite{Park03,Pastore08,Pastore09,Kolling09,Kolling11}.

The $\chi$EFT procedure briefly outlined above has a few advantages over a pure phenomenological approach: 
\begin{itemize}
\item An expansion in $(Q/\Lambda_\chi)^\nu$ allows
 to evaluate nuclear observables to any degree $\nu$ of desired accuracy, with an
associated theoretical error, which can be roughly estimated by $(Q/\Lambda_\chi)^{(\nu+1)}$. Thus, this approach is systematic 
and allows to estimate theoretical error bars, at least in principle;\\

\item  NN forces, such as $V_{ij}$ in Eq.~(\ref{pot}), and  3N forces, such as the $W_{ijk}$ in Eq.~(\ref{pot}), appear naturally and consistently with each other. 3N forces are sub-leading with respect to the NN forces, as, in the considered scheme, they first appear at N2LO.\\

\item Current operators that couple  nucleons to external electroweak probes can be in principle constructed consistently with the potentials at different orders.

\end{itemize}

Since the pioneering work of Weinberg~\cite{Weinberg90},
$\chi$EFT has been widely used in nuclear physics and
 has developed into an intense field of research. Several applications appeared in the sector of light-nuclei, see, {\it e.g.}, Refs.~\cite{Epelbaum12,Machleidt11}, but also heavier systems are being studied~\cite{Hebeler2015}.

In particular, the strategy of $\chi$EFT offers  a tool to derive forces that can be utilized in interacting shell model calculations, as opposed to using potentials which are derived by phenomenologically fitting matrix elements. This approach has been in fact successfully used to address some physics cases of interest to the field of rare isotopes. A description of two specific examples regarding the oxygen and the calcium isotopic chains will be presented in the next section. Before that,  we will introduce ab initio methods, which also make use of  $\chi$EFT interactions and aim at a full solution of the many-body problem.

\section{Ab initio methods}
The phenomenological interacting  shell model has provided us with deep insight into nuclear structure.  However, intrinsic into the definition of its scheme is the  {\it ansatz}  of a core, a valence and an external space. These assumptions make it difficult to quantify the theoretical error bars intrinsic to the model. 
Recently, enormous progress has been done to develop many-body methods which go beyond such approximations. Ideally, one would like to be able to solve the quantum problem of many nucleons interacting with each other, as described by Eq.~(\ref{schro}), without introducing approximations.
Ab initio methods usually refer to computational techniques to solve Eq.~(\ref{schro})
in a numerically exact way or within controlled approximation schemes, which give the possibility to assess theoretical error bars. 

Today, a number of methods exists, which can be included in this category, see, {\it e.g.},~\cite{Leidemann2013,Bacca2014,Hebeler2015} for recent reviews.
Methods commonly utilized for $A\leq4$ systems include 
the Faddeev-Yakubovsky scheme~\cite{Yakubovsky66},
the  hyperspherical harmonics method~\cite{Viviani05,Kievsky08} and the no-core shell model (NCSM)~\cite{Barrett13,navratil2009}.
Other powerful ab initio methods that can tackle nuclei with mass number $A>4$ (with different applicability  when the mass range is augmented) are 
the effective interaction hyperspherical harmonics expansion \cite{Barnea00,barnea2001}, Green's function Monte Carlo  methods \cite{pieper2001,Lovato}, the auxiliary field diffusion Monte Carlo method~\cite{AFDMC}, the NCSM also when used in conjunction with truncation schemes \cite{IT-NCSM}, coupled-cluster methods \cite{hagen2010b,CC_review}, the 
fermionic molecular dynamics  approach \cite{Neff2008}, 
in medium similarity renormalization group \cite{ImSRG,MR-IM-SRG}, 
self--consistent Green's function theory \cite{carlo} and lattice simulations \cite{latticeEFT}.

 While presenting an exhaustive and complete description of all ab initio methods goes beyond the scope of these notes, this Section contains a brief introduction to diagonalization methods and coupled-cluster theory.
Subsequently, the Lorentz integral transform method  will be introduced as an ab initio way to solve the multi-channel continuum problem in nuclear reactions induced by perturbative probes.
 Some space is  devoted to the presentation of applications involving rare isotopes where ab initio theoretical results are compared to experiment. For organization purposes, such applications are embedded in the Section in the form of paragraphs.

\subsection{Diagonalization methods}

Among all ab initio methods, some can be easily understood using  concepts that were introduced in  Section~\ref{ism} on the interacting shell model.
For example, one could solve the Schr{\"o}dinger equation in (\ref{schro}) by expanding the many-body state in terms of a complete set of basis states, similarly to what expressed in Eq.~(\ref{exp}). In particular, if one used  harmonic oscillator single-particle states and considered a many-body space spanned by Slater determinants obtained with particle-hole excitations in all orbitals, one would basically perform a NCSM calculation. In essence, this would mean considering all orbitals as 
 valence space, as opposed to introducing a core and an external space as done in Figure~\ref{gabriel}. It is straightforward to see that the problem is reduced to the diagonalization of a Hamiltonian matrix represented on that specific set of basis states.  Hence, these techniques are named diagonalization methods.

Typically, in ab initio calculations translational invariant operators are used, {\it i.e.}, operators which do not depend on the motion of the center of mass. Thus, the diagonalized Hamiltonian is
\begin{equation}
H=K-K_{\rm CoM}+\sum_{i<j}^AV_{ij}+\sum_{i<j<k}^AW_{ijk}+ \dots \,, 
\label{hamilt3}
\end{equation}
where $K_{\rm CoM}$ is the kinetic energy of the center of mass of the nucleus
\begin{equation}
K_{\rm CoM} = \sum_i^A \frac{p^2_i}{2mA}\,.
\end{equation}
Notice that  NN and 3N potentials typically depend only on relative coordinates, not on the center of mass.
NCSM is capable of utilizing NN and 3N forces from $\chi$EFT and has been applied to a variety of nuclei,
see {\it e.g.} Refs.~\cite{Barrett13,navratil2009} for recent reviews. 

\begin{figure}[t]
\begin{center}
\includegraphics[width=0.19\textwidth,clip=]{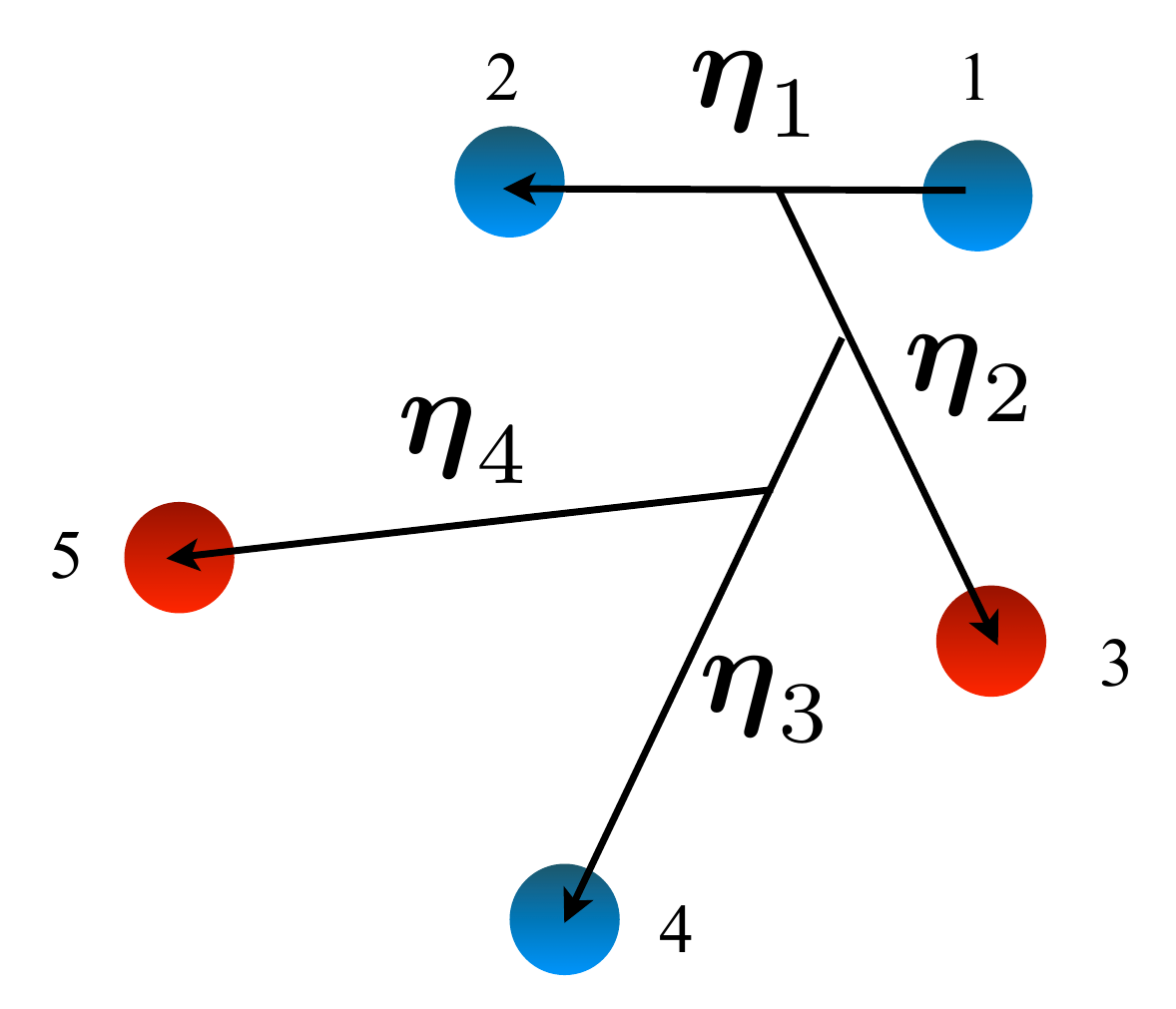}
\caption{Example of Jacobi coordinates for a system of $A=5$ nucleons.}
\label{jac}
\end{center}
\end{figure}
Another ab initio method which is based on the expansion of the many-body wave function on a particular set of basis states, and thus belongs to the category of  diagonalization methods, is the hyperspherical harmonics technique. First, instead of working with laboratory coordinates  $\left\{{\bf r}_i \right\}$ one introduces relative coordinates for each particle
\begin{equation}
\mathbf{r}_{i}^{\prime }=\mathbf{r}_{i}-\mathbf{R}_{\rm CoM}, 
\label{transf} 
\end{equation}
where
\[
{\bf R}_{\rm CoM}=\frac{1}{A}\sum ^{A}_{i=1}{\bf r}_{i}\]
denotes the center of mass coordinate.  The goal is then to work  in the  center of mass frame, where one has
\[\sum_i^A {\bf r}'_i=0.
\] 
Because of the form of the Hamiltonian in Eq.~(\ref{hamilt3}), the total many-body wave function can be factorized in center of mass and internal parts as
$\Psi=\Psi_{\rm CoM}~\Psi_{int}$. 
The internal wave function can be described  in terms of  a set of $A-1$  independent 3-dimensional Jacobi coordinates, $\{\boldsymbol {\eta }_{k}, k=1,...,A-1 \}$, defined as
\begin{equation}
\boldsymbol {\eta }_{k-1}=\sqrt{\frac{k-1}{k}}\left( \mathbf{r}_{k}-\frac{1}{k-1}\sum _{i=1}^{k-1}\mathbf{r}_{i}\right) ;\, \, \, k=2,...,A.
\label{jacobc}
\end{equation}
As one can read from Eq.~(\ref{jacobc}), Jacobi coordinates are basically proportional to the relative distance between the $k-$th particle coordinate and the center of mass of the remaining $(k-1)$-body system.
Instead of using the $\{{\bf r}'_i\}$ coordinates, which are not linearly independent, one can use the $\{\boldsymbol{\eta}_k\}$ coordinates.
 Figure~\ref{jac} shows a diagrammatic representation of the Jacobi coordinates for mass number $A=5$. Jacobi coordinates are also used in NCSM calculations for nuclei with mass number $A\le 5$.

\begin{figure}[t]
\begin{center}
\includegraphics[width=0.46\textwidth,clip=]{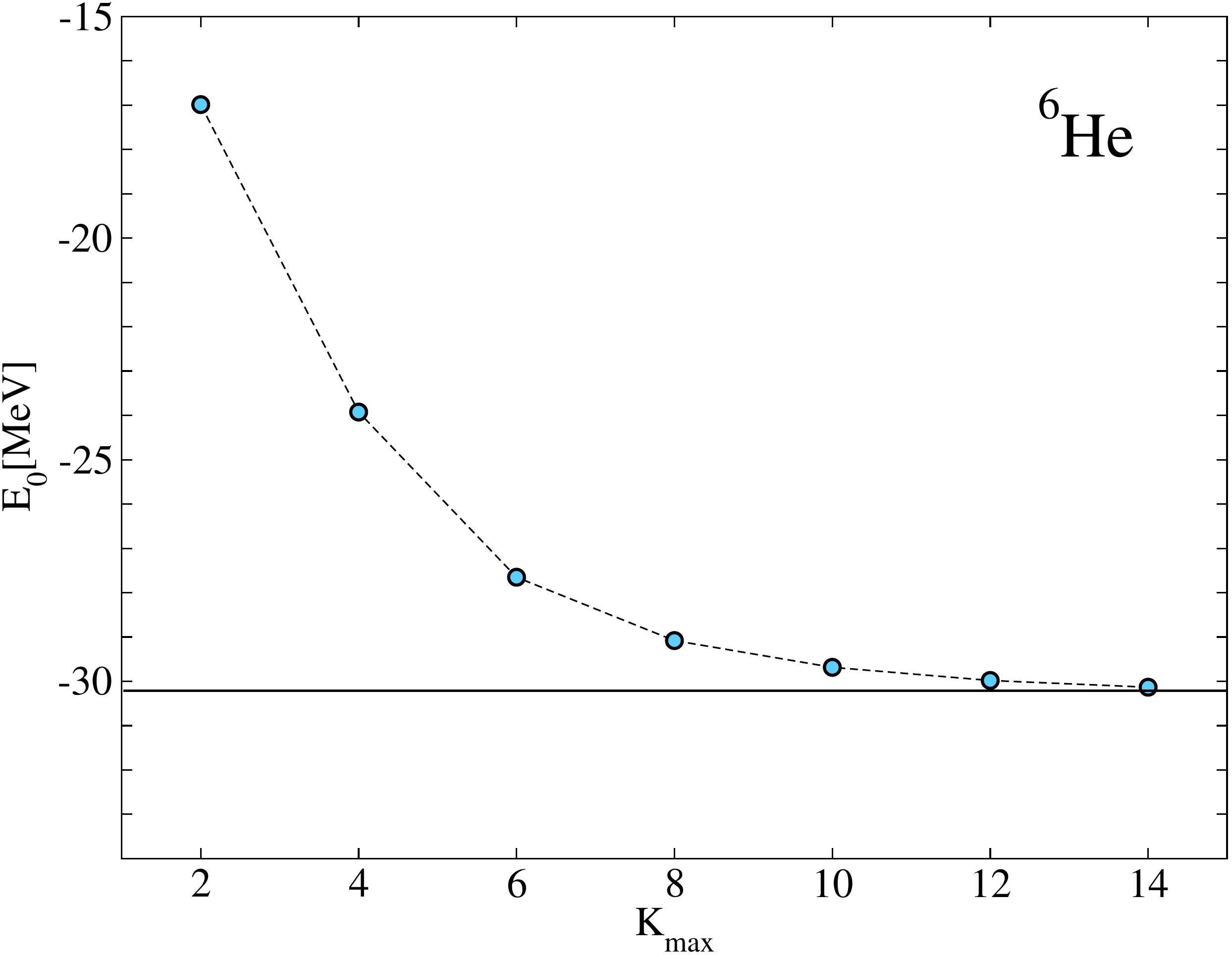}
\caption{Convergence of $^6$He binding energy as a function of the maximal grandangular momentum $K_{max}$ for a two-body soft interaction. See details in Ref.~\cite{EIHH}. Similar plots are obtained for other light nuclei.}
\label{he6conv}
\end{center}
\end{figure}

Starting from the Jacobi coordinates one can apply the recursive transformation to obtain hyperspherical coordinates as
\begin{eqnarray}
\rho _{k-1} & = & \rho _{k}\cos \varphi _{k},\nonumber \\
\eta _{k} & = & \rho _{k}\sin \varphi _{k},\label{ipN} 
\end{eqnarray}
 where \begin{equation}
\label{iperrad}
\rho _{k}^{2}=\rho _{k-1}^{2}+\eta _{k}^{2}=\sum ^{k}_{i=1}\eta _{i}^{2}=\frac{1}{k}\sum ^{k+1}_{i<j}\left( \mathbf{r}_{i}-\mathbf{r}_{j}\right) ^{2}
\end{equation}
is the hyperradius and \( \varphi _{k} \) are the hyperangles.
The total of \( 3\left( A-1\right)  \) internal coordinates are then redefined
 in terms of one hyperradial coordinate
\( \rho _{A} \),   \( A-2 \) hyperangular coordinates
\(\left\{ \varphi _{2},\varphi _{3},...,\varphi _{A-1}\right\},  \)
and finally by \( 2\left( A-1\right)  \) angular coordinates coming from the Jacobi vectors \(\left\{ \hat{\eta}_{i}\right\}  \).
For simplicity we denote all the angular coordinates with  a collective symbol \( \Omega \).
These coordinates depend on the set of starting Jacobi coordinates, since when changing the indices of the particles a different set of hyperangular and angular coordinates is obtained. Only the hyperradius remains unchanged.

After making the transition to hyperspherical coordinates, one expands the  $A$-body internal wave functions $\Psi_{int}$ in terms of 
 hyperspherical harmonics and hyperradial functions as
\begin{equation}
\Psi_{int}( \boldsymbol {\eta }_{1}, ..., \boldsymbol {\eta }_{A-1})=\sum_{K \nu}^{K_{max}\nu_{max}} c_{K \nu}~ R_{\nu}(\rho){\mathcal H}_{K}(\Omega)\,,
\label{expans}
\end{equation}
where for simplicity we just write spatial coordinates and not spin or isospin degrees of freedom.
The hyperspherical harmonics $\mathcal H_{K}$ for an $A$-body system are labeled by the quantum number $K$ which is called grandangular momentum, while hyperradial states are labeled by a quantum number $\nu$.
Expansions have to be performed up to maximal values of such quantum numbers, $K_{max}$ and $\nu_{max}$, respectively. Hyperspherical harmonics ${\mathcal H}_{K}$ 
are  linear combinations of products of Jacobi polynomials times ordinary spherical harmonics~\cite{key-25}. 
They do not possess any peculiar property under permutation of particles.
Thus, since we are working with identical fermions it is convenient for the basis states to be fully antisymmetrized. Details on the antisymmetrization procedure are beyond the scope of these lectures.  The interested reader can find information on one possible way to antisymmetrize hyperspherical harmonics in Refs.~\cite{BN97,BN98}. The advantages of the hyperspherical harmonic basis are that:
\begin{itemize}
\item It avoids center of mass problems, by expanding directly only internal wave functions;\\

\item It converges rather fast as a function of $K_{max}$ and $\nu_{max}$;\\

\item The hyperradial functions $R_{\nu}(\rho)$ can be chosen so that they have an exponential fall-off, thus representing the correct asymptotic behavior of the wave functions, as opposed to the Gaussian fall-off imposed by the harmonic oscillator basis. 
\end{itemize}

\begin{figure}[t]
\begin{center}
\includegraphics[width=0.48\textwidth,clip=]{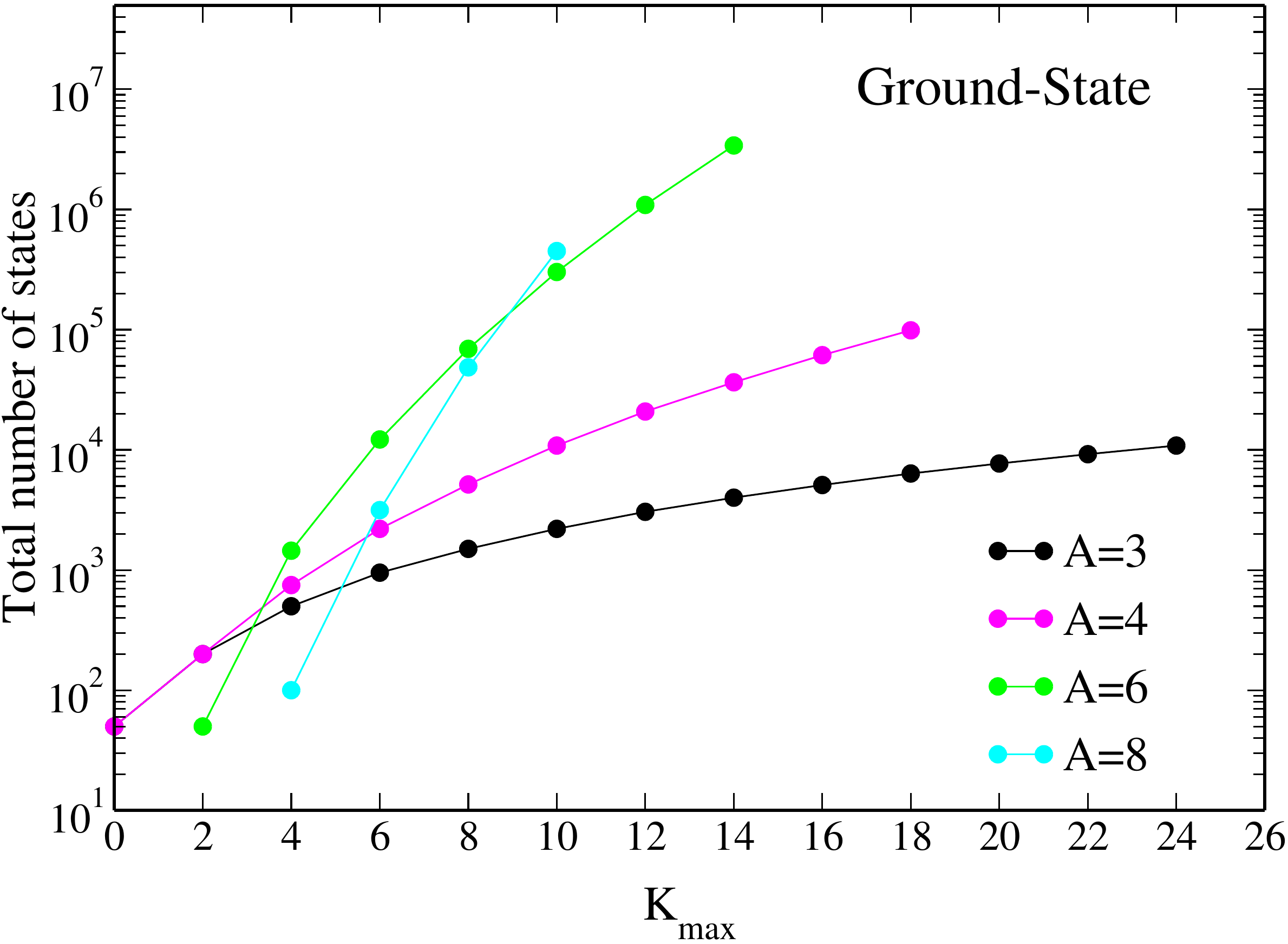}
\caption{Total number of antisymmetrized basis states in a hyperspherical harmonics expansion as a function of $K_{max}$ for different mass numbers. The number of hyperradial states is kept constant to 50.}
\label{scaling}
\end{center}
\end{figure}

A crucial aspect of hyperspherical harmonics method is that the expansion in Eq.~(\ref{expans}) has to be carried out up to the virtual infinity, {\it i.e.}, up to values of the maximal grandangular momentum $K_{max}$ and maximal number of hyperradial states $\nu_{max}$  where calculated observables do not depend anymore on these truncations.
In Figure~\ref{he6conv}, the example of the $^6$He nucleus is shown, calculated with a soft two-body force~\cite{EIHH}.  While the convergence in terms of $\nu_{max}$ can be reached with about 50 states or less, the convergence in  $K_{max}$ is typically a lot more delicate and needs to be carefully investigated. From Figure~\ref{he6conv} it is clear that once $K_{max}=12$ or 14 is reached, the ground-state energy $E_0=-BE(^6{\rm He})$ is independent of the parameter $K_{max}$. 
As  $K_{max}$ grows, however, the number of basis states increases dramatically. This means that the dimensionality of the dense matrix to diagonalize becomes very big, as shown  in Figure~\ref{scaling}. When the matrix dimension becomes of the order of $10^{6}$ or $10^7$,  the problem is not tractable anymore. That is one of the reasons why direct diagonalization methods are limited in mass number. 

It is worth mentioning that in the NCSM approach, when a Slater determinant basis is used, then much larger matrices can be diagonalized, as they turn out to be sparse  and not dense, {\it i.e.}, with many null matrix elements. Nevertheless, when the mass number $A$ and model space size increase, the dimension explodes. When strategies are identified on how to discard unimportant states from the expansion, then larger system can be investigated. This is the case, for example with the importance-truncation no-core shell model, see, {\it e.g.}, Ref.~\cite{IT-NCSM}.

\paragraph{The halo nucleus $^6$He }
\begin{figure}[t]
\begin{center}
\includegraphics[width=0.19\textwidth,clip=]{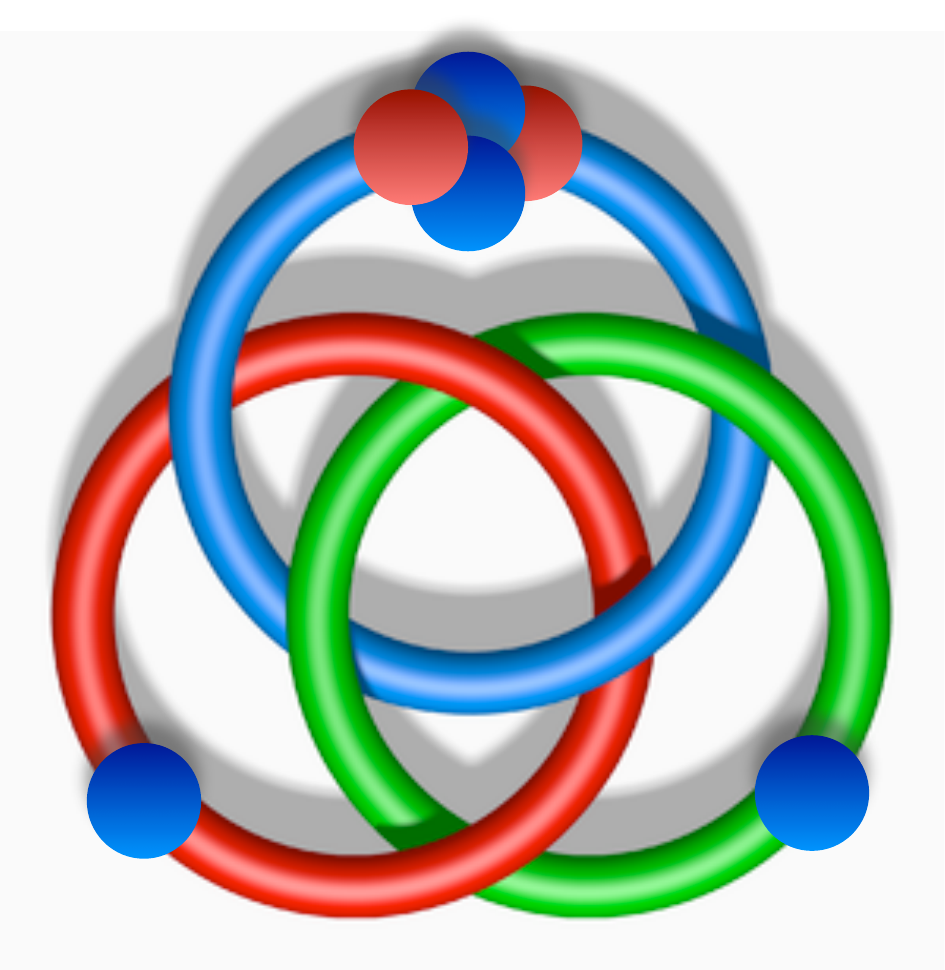}
\caption{Borromean rings as a schematization of borromean nuclei, such as $^6$He, composed by a $^4$He core and two neutrons held together by the strong force.
\label{fig:borro}}
\end{center}
\end{figure}

As an example of ab initio calculations in light nuclei  relevant to the physics of rare isotope, the case of the 
$^6$He nucleus will be presented next and compared to experiment.

$^6$He is a halo nucleus which undergoes $\beta$ decay and has a half life of about 0.8 seconds. 
Halo nuclei are exotic structures that arise in the sector of light nuclei, when there is an excess of one nucleon species with respect to the other. In such cases, the excess nucleons orbit away from all the others, forming a kind of ``halo''. With 4 neutrons and 2 protons, $^6$He is a two-neutron halo system where the halo neutrons are bound to the $^4$He core by roughly 1 MeV. In particular, $^6$He is also a borromean nucleus, in the sense that, if viewed as a three-body system composed by the core and two neutrons, none of the two-body sub-systems
is bound~\cite{jonson}, similarly to what happens with the borromean rings  shown in Figure~\ref{fig:borro}.

 What characterizes neutron halo nuclei are very small neutron separation energies and large matter radii, compared to  charge radii,
indicating that neutrons are much more diffused than protons.
 Ab initio calculations of halo nuclei are  challenging, precisely due to the fact that wave functions are very extended.
Nonetheless, several calculations have been performed  for $^6$He and compared to experimental data. In fact,
despite the short half life, precise measurements of masses and charge radii can be performed by trapping exotic ions
produced at the rare isotope facilities.
\begin{figure}[t]
\begin{center}
\includegraphics[width=0.53\textwidth,clip=]{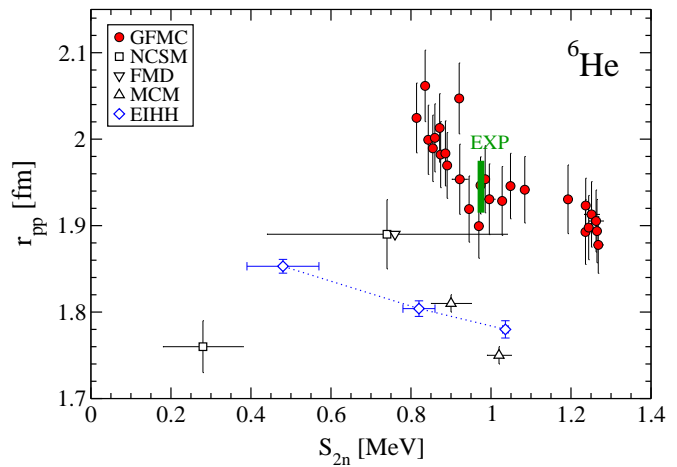}
\caption{The $^6$He point-proton
radius $r_{\rm pp}$ versus two-neutron separation energy
$S_{2n}$. The experimental range (bar) is compared to theory based
on different ab initio calculations where different NN interactions 
are used (open symbols) and where  3N forces are included (filled red
symbols).  More details
can be found in Ref.~\cite{Brodeur}.
\label{fig:He6_Sn_rpp}}
\end{center}
\end{figure}
The binding energy of $^6$He was recently measured very precisely at TRIUMF with the TITAN Penning trap~\cite{Brodeur}
and its value was then used to extract the charge radius from previous laser spectroscopy measurements~\cite{Wan04}, 
 using sophisticated theoretical calculations in atomic physics.
The experimental results, together with all the available ab initio calculations are shown in Figure~\ref{fig:He6_Sn_rpp}.

The measured charge radius is converted to a point-proton radius $r_{\rm pp}$  using the following relation~\cite{Ong10}
\begin{equation}
r^{2}_{\rm pp} = r^{2}_{\rm c} - R^{2}_{p} - (N/Z) \cdot R^{2}_{n} -
3/(4 M^{2}_{p}) - r^{2}_{\rm so}\,,
\label{rpp}
\end{equation}
where $R^{2}_{p}$ and $R^{2}_{n} = -0.1161(22)$ fm$^{2}$
are the proton and neutron mean-square charge radii, respectively.
The quantity $3/(4 M^{2}_{p}) = 0.033$ fm$^{2}$ is a  relativistic
correction~\cite{Fri97} named Darwin-Foldy and $r^{2}_{\rm so}$ is a
spin-orbit nuclear charge-density correction.
For $R_{p}$  the combined values of  0.877(7)fm~\cite{Nak10} and  $0.84184(67)$~fm~\cite{Poh10} were used, while for
the spin-orbit radius the coarse mean field $0.17$~fm estimate by Ref.~\cite{Ong10} was utilized
with the same value as an error bar. This has lead to the range of variation of $r_{\rm pp}$
in Figure~\ref{fig:He6_Sn_rpp}. The separation energy, being the difference between two binding energies,  is instead measured very accurately and there is virtually no spread in $S_{2n}$ in  Figure~\ref{fig:He6_Sn_rpp}.

Several ab initio calculations are also shown in Figure~\ref{fig:He6_Sn_rpp}.
The Green's
Function Monte Carlo (GFMC) results~\cite{GFMC} are the only existing
converged calculations that include 3N forces, which are calibrated
on properties of light nuclei, including $^6$He. The spread in the points
gives an idea of the uncertainties involved in the calculations, especially
in the 3N force models used (two-different models were adopted, see Ref.~\cite{GFMC}
for more details). 
All the other empty symbols correspond to calculations performed with NN forces only.
In particular, the diamonds correspond to effective interaction hyperspherical harmonics (EIHH) calculations from Refs. \cite{Brodeur,EIHH} 
based on chiral low-momentum NN
interactions $V_{{\rm low}\,k}$~\cite{forces}. By varying the cutoff of the underlying nuclear force, which is a
 degree of freedom, one observes an expected correlation between the two observables: as the separation energy 
approaches zero, the radius becomes larger as the system becomes unbound. 
Other calculations include
 the fermionic molecular dynamics (FMD)
result \cite{FMD}, the
NCSM results \cite{NCSM}, and the variational microscopic cluster model
(MCM) results~\cite{MCM}. 

Because the only calculations that go through the experimental band include 3N forces,
Figure~\ref{fig:He6_Sn_rpp} shows their importance in the physics or rare isotopes.
Finally, it is worth mentioning that novel ab initio approaches are being presently developed to study $^6$He using the NCSM and augmenting it by providing the correct long range behavior of the wave functions~\cite{carolina1,carolina2}.

\subsection{Coupled-cluster theory}

An alternative way to solve the many-body problem is provided by coupled-cluster theory, which was introduced originally by Coester and K\"{u}mmel~\cite{coesterkummel}. Coupled-cluster theory is being widely used in chemistry~\cite{shavittbartlett2009} and has recently had a renaissance in nuclear physics, see, {\it e.g.}, the recent review~\cite{CC_review} and references therein.

By starting from a reference Slater determinant $\Phi_0$, this theory assumes that the full many-body ground-state can be found using the following exponential {\it ansatz}
\begin{equation}
 \Psi= e^{T}\Phi_0 \,.
\end{equation}
Here, the operator $T$ is a correlation operator and can be expanded into particle-hole ($p-h$) excitation operators as
\begin{equation}
T=T_1+T_2+T_3+\dots \,,
\end{equation}
where $T_1$ is a one-particle one-hole ($1p-1h$) excitation operator, $T_2$ is a two-particle two-hole excitation  ($2p-2h$) operator, $T_3$ is a three-particle three-hole excitation ($3p-3h$), etc.
Using the second-order quantization language these operators can be written as
\begin{eqnarray}
\nonumber
T_1&=&\sum_{ia} t_{i}^a a_a^\dagger a_i,\\
\nonumber
T_2&=&\frac{1}{4}\sum_{ijab} t_{ij}^{ab} a_a^\dagger a_b^\dagger a_j a_i,\\
T_3&=&\frac{1}{36}\sum_{ijkabc} t_{ijk}^{abc} a_a^\dagger a_b^\dagger a_c^\dagger a_k a_j a_i \,,\\
\cdots & &\cdots
\label{top}
\end{eqnarray}
 where indexes $i, j, k,\dots $  indicate occupied single--particle (hole) states in the reference Slater determinant, while the $a, b, c, \dots $  label unoccupied (particle) states.
 
Next, one introduces the similarity transformed
Hamiltonian 
\begin{equation}\label{hbar}
 \overline{H} = \exp(-T)  H \exp(T).
\end{equation}
In terms of the similarity transformed Hamiltonian, the many-body Schr\"{o}dinger equation for the ground-state becomes
\begin{equation}
\overline{H} \Phi_0 = E_0 \Phi_0 \,.
\end{equation}
The amplitudes of the $T$ operator, such as $t_i^a$ , $t_{ij}^{ab}$, $t_{ijk}^{abc}$, etc.,  can be found by solving the set of non-linear equations given by \begin{eqnarray}
\nonumber
0&=&\langle \Phi_i^a| {\overline H}| \Phi_0 \rangle,\\
0&=& \langle \Phi_{ij}^{ab}| {\overline H}| \Phi_0 \rangle,\\
\nonumber
0&=& \langle \Phi_{ijk}^{abc}| {\overline H}| \Phi_0 \rangle,\\
\nonumber
 &&\cdots \,.
\end{eqnarray}
Here $\Phi_{i}^a $, $\Phi_{ij}^{ab} $ and $\Phi_{ijk}^{abc}$ are Slater determinants constructed as $1p-1h$, $2p-2h$, $3p-3h$, $\dots$ excitations  on top of the reference state, respectively.  More detailed information can be found in Refs.~\cite{CC_review,shavittbartlett2009}.

This theory is exact when the expansion of the $T$ operator is performed up to $Ap-Ah$ excitations. However, due to the exponential {\it ansatz}, even when truncations schemes are introduced, the result is much closer to the exact one with respect to when a linear {\it ansatz} is done, as in the diagonalization methods.
For example, when the expansion in $T$ is truncated at the $2p-2h$ level, named coupled-cluster with single and double (CCSD) excitation, only about $10\%$ of the correlation energy is missed with chiral potentials such as \cite{Entem03}.
 When approximate triples are added, almost all correlations are included~\cite{Gaute_triples}. 
The advantage of the method is that it scales mildly with increasing mass number, with the computational load  behaving with a polynomial law, thus not exponentially fast as shown before in Figure~\ref{scaling} for diagonalization methods.
Thus, this approach is very powerful.

Coupled-cluster theory in the presented single-reference formulation is applicable to double magic nuclei or nuclei near magic numbers.
With equation of motion methods~\cite{jansen} ground- and excited-states can be calculated for closed sub-shell nuclei and one- or two-particle attached or removed systems from the closed-sub-shell nucleus.
This powerful theory has been applied to a variety of systems and we refer the interested reader to the recent review~\cite{CC_review} for an update on applications.

\paragraph{Neutron drip line in  oxygen and calcium isotope chains}
As an application of coupled-cluster theory and other ab initio theories relevant to the physics of rare isotopes  we present below  studies of the neutron drip line in the oxygen and calcium  chains.

One of the central challenges in structure studies for exotic nuclei is to understand and predict
the location of the neutron drip line. The latter is the point where nuclei cease to be bound and neutron separation energy becomes zero. In other words, if we start to pack neutrons on a given element with $Z$ protons, at some point in the $N>Z$ regime the last neutron will fall apart. This point in neutron number determines the neutron drip line. The latter is experimentally known only for light nuclei and more is to be discovered in the future
concerning heavier systems at the rare isotope facilities. From the theoretical point of view, studies have shown increasing evidence of details of the Hamiltonian and 3N forces being very important in determining where nuclei stop being bound, see, {\it e.g.}, Ref.~\cite{Hebeler2015}.

 In particular, the oxygen isotopic chain is very interesting, because the neutron drip line found with $N=16$ is anomalously close to the line of stability, only 6 neutrons away. Moreover, there are three bound nuclei of closed shell nature, $^{16}$O, $^{22}$O and $^{24}$O, with the latter being located exactly at the drip line, making it very amenable to coupled-cluster theory. 

 Interacting shell model calculations with phenomenological potentials predict that $^{24}$O is the last bound isotope.
However, if one were to adopt the modern approach of $\chi$EFT and used only NN soft forces~\cite{forces}, one would obtain that oxygen isotopes heavier than  $^{24}$O are bound, opposite to observation. Insight  was gained into the mechanism that determines the location of the drip line once 3N forces were introduced in 
shell model calculations~\cite{Otsuka2010}. By using 3N forces represented by diagrams in Figure~\ref{3nf}, and considering the case where two nucleons are in the valence space and one in the core, it was possible to determine that 3N forces are essential to explain that $^{24}$O is the last bound oxygen isotope. This is a very nice example where the shell model scheme was used together with modern realistic forces to shed light on phenomenology.
\begin{figure}
\centering
  \includegraphics[width=\linewidth]{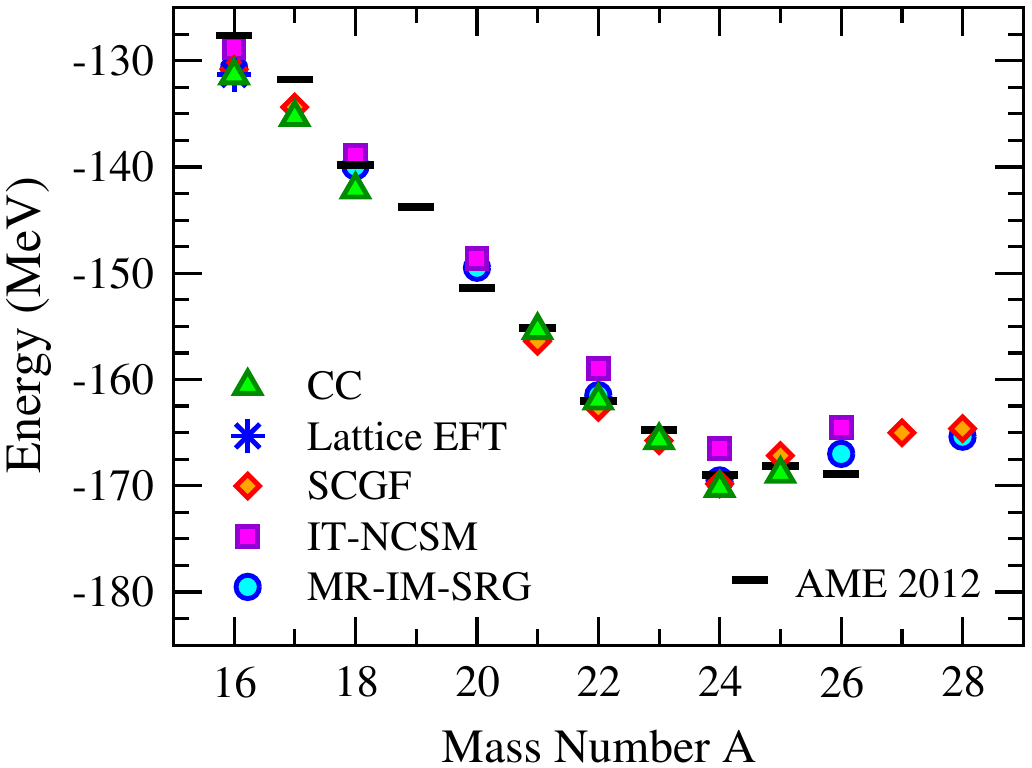}
  \caption{Drip line in the oxygen isotope chain: results from different ab initio methods employing Hamiltonians derived in chiral effective field theory and including three nucleon forces. Theoretical results are compared to experimental data from the atomic mass evaluation of 2012.
Figure adapted from Ref.~\cite{Hebeler2015}.}
 \label{bench_fig}
\end{figure}

Shell model calculations are however limited by the choice of a core and model space, thus they are not completely free of approximations.
 More recently, several ab initio approaches were able to address the very same systems, going beyond the core-approximation. This provided a good platform for benchmarking different methods, giving an opportunity to assess the accuracy of numerical simulations.

In Figure~\ref{bench_fig}, ground-state energy of oxygen isotopes are shown versus mass number. Several different calculations are compared with one another. Besides coupled-cluster theory labeled  with CC, where triples are included in a non-perturbative but approximate way, other methods shown include multi--reference in medium similarity renormalization group (MR-IM-SRG)~\cite{MR-IM-SRG},  self--consistent Green's function (SCGF) theory~\cite{carlo}, the importance--truncation no--core shell model (IT-NCSM)~\cite{IT-NCSM}, which extends an exact NCSM diagonalization by sampling important states, and finally, nuclear lattice effective field theory  (lattice EFT) simulations for $^{16}$O~\cite{latticeEFT}. 

Besides the details of everyone of these ab initio methods, whose explanation goes beyond the scope of these notes, it is interesting to know that, when the same Hamiltonian is used, different numerical simulations provide very close results.
Here, besides the lattice EFT results, all other methods employed an evolved chiral interaction~\cite{SRG1,SRG2}, which   includes two-body forces~\cite{Entem03} and  3N forces~\cite{Gazit09}.

It is clear that all methods  predict the location of the drip line to be at the $^{24}$O nucleus. As in case of the shell model calculations, all these ab initio methods confirm  that in the absence of 3N forces, isotopes heavier than $^{24}$O would be bound, in contradiction to experimental observations.  Thus, 3N forces are essential in describing the correct location of the drip line.
The small variation that one observes in Figure~\ref{bench_fig} between the different methods can be interpreted as an  estimate of the overall accuracy we can reach today with modern ab initio methods, excluding the indetermination one has from the use of different Hamiltonians.

\begin{figure}
\centering
  \includegraphics[width=0.95\linewidth]{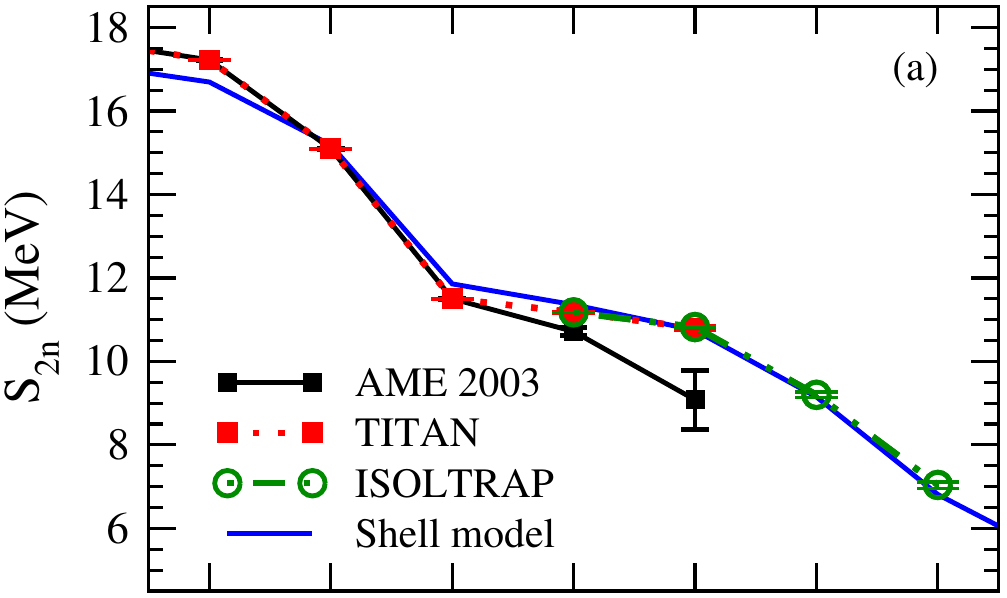}
  \includegraphics[width=0.95\linewidth]{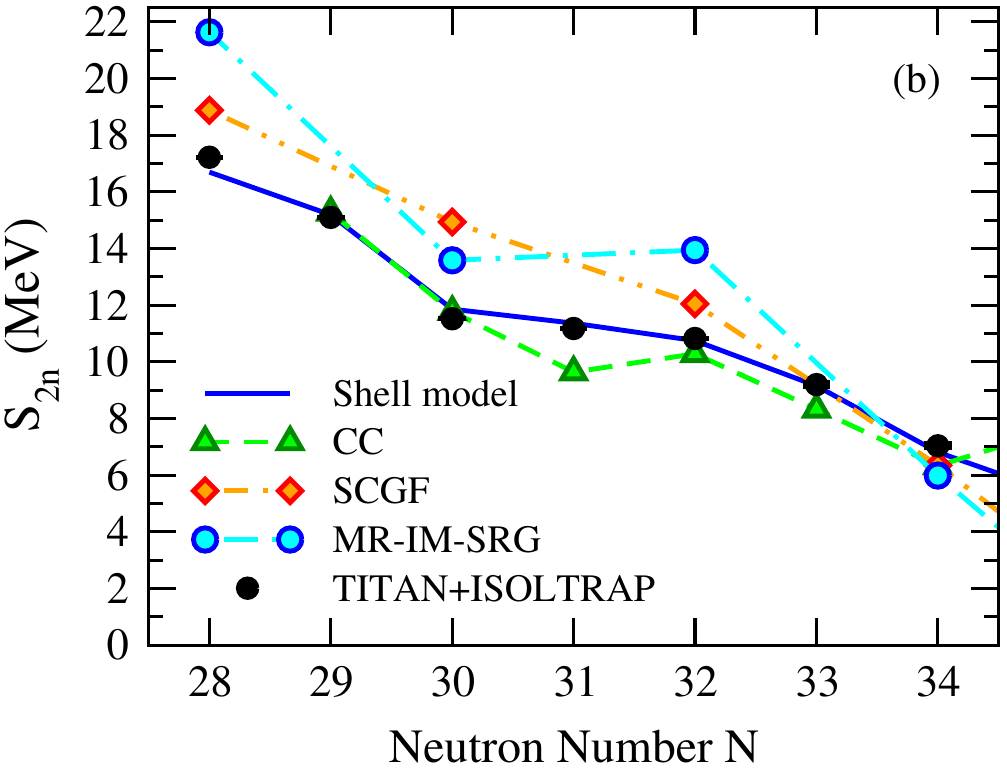}
  \caption{Two-neutron separation energy (difference of binding energies) of neutron-rich Ca isotopes as a function of neutron number $N$. Panel (a): Measurements by TITAN and ISOLTRAP in comparison to the atomic mass evaluations of 2003 and interacting shell model calculations with 3N forces (blue line). Panel (b): All available ab initio calculations which include 3N forces (slightly different ones) in comparison to the experimental data by TITAN and ISOLTRAP. Figures adapted from Refs.~\cite{Hebeler2015} and \cite{nature}.
\label{calcium_s2n}}
\end{figure}

Another isotopic chain that was recently investigated both theoretically and experimentally is the calcium one.
 Atomic mass measurements performed with the TITAN Penning trap at TRIUMF  provided precise access to the nuclear binding energies of Ca isotopes~\cite{gallant}. This allowed to make interesting comparison to calculations~\cite{Holt2012,Coraggio3,Coraggio4} based on shell model with a $^{40}$Ca core and where 3N forces from $\chi$EFT were included with the assumption that one interacting nucleon is in the core. Figure~\ref{calcium_s2n}(a) shows such a comparison in case of the two-neutron separation energy.  The TRIUMF experimental results found that $^{52}$Ca deviated by almost 2 MeV from the previous atomic mass evaluation (AME 2003), but agrees well with the predictions from shell model calculations  with 3N forces. 
More recently, the ISOLTRAP collaboration at ISOLDE/CERN confirmed the TITAN results and was able to 
further advance the limits of precision mass measurements, reaching out to the more neutron-rich  $^{53}$Ca and $^{54}$Ca nuclei. Both  $^{53,54}$Ca measurements are in excellent agreement with predictions from  shell model theory with modern potentials and also establish N = 32 as a shell closure~\cite{nature}. 

 Figure~\ref{calcium_s2n}(b) shows, instead, all available ab initio calculations  which include 3N forces in comparison to the experimental data measured by TITAN and ISOLTRAP. Theoretical results include, on top of the shell model calculations from Ref.~\cite{gallant}, the CC~\cite{Hagen_ca}, the SCGF~\cite{soma} and the MR-IM-SRG~\cite{heiko} results, with acronyms as introduced above. In this case, each computations used 3N force models derived in $\chi$EFT, but with slightly different parameterizations. Thus, the difference observed in the various results has not to be interpreted solely as the numerical error of the many-body methods, but rather as an estimate of the ``nuclear physics'' error, due to the fact that we do not have one model for the nuclear force, but several parameterization are justified. 

Finally, thanks to the enormous progress of theoretical computations, neutron drip line can be investigated with ab initio methods and some uncertainties can be quantified. Evidently, more work needs to be done to reduce theoretical error bars, since  they are still much larger than the experimental ones.

\subsection{Lorentz integral transform method}

\begin{figure*}[t]
\centering
  \includegraphics[width=0.8\textwidth]{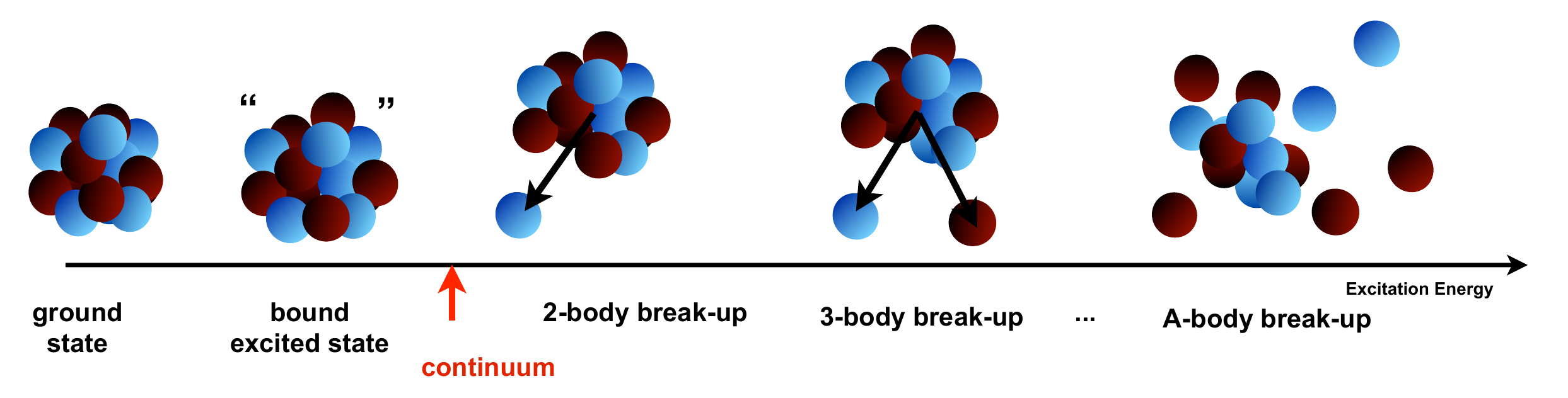}
  \caption{Schematic representation of the spectrum of a nucleus. Above particle emission threshold, typically first the two-body break-up channel opens up, then the three-body break-up channel up to the $A$-body break-up channel.}
\label{cont_fig}
\end{figure*}

While ground-state properties in nuclei are among the most important observables investigated in nuclear physics,
excited states and inelastic processes  allow  to study further aspects of nuclear dynamics.
In particular, interactions of the nucleus with electroweak probes are very important because the contribution of the external probe can be disentangled from the dynamics of the strong force. Furthermore,   
the perturbative nature of the process allows to clearly  connect measured cross sections with the calculated structure properties of nuclear targets.

Typically, electroweak cross sections depend on the nuclear response function
\begin{equation} 
R(\omega)=\sum_n \left|\langle \Psi_n|{\mathcal O}|\Psi_0 \rangle \right|^2\delta\left(E_n-E_0-\omega \right),
\label{eq:rs}
\end{equation}
where $\mathcal O$ is the excitation operator, which will depend specifically on the external probe and $\Psi_n$ are the excited states of the nucleus.
Thus, the nuclear response function is a dynamical observables which requires knowledge of the whole spectrum of the nucleus. This includes not only bound excited-states, but also excited states in the continuum above particle emission threshold $\omega_{th}$, see Figure~\ref{cont_fig}.

Despite the enormous progress 
achieved in the ab initio calculation of ground-state energy and spectra of nuclei with increasing mass numbers,
the  exact calculation of final state many-body continuum wave functions constitutes  
still an open theoretical problem.
Most of the  ab initio studies of electroweak break-up reactions are performed for 
systems with  $A \lesssim 4$ and at energies below the
three-body break-up threshold.
The difficulty in calculating a many-body cross section 
involving continuum states is that at a given energy
the wave function of the system can have different channels
corresponding to all its partitions into fragments of various sizes, see Figure~\ref{cont_fig}.
In particular,  the implementation of the boundary conditions
for a wave function in the continuum  constitutes the main obstacle. 
However, integral transform approaches allow  
 to reformulate the problem so that knowledge of the continuum states is not necessary.

In particular,  the Lorentz integral transform (LIT) of the response function~\cite{Efl07}
is defined as
\begin{equation}
  { L}(\omega_0,\Gamma )=\frac{\Gamma}{\pi}\int_{\omega_{th}}^{\infty} d\omega \frac{R(\omega)}{(\omega -\omega_0)
               ^2+\Gamma^2}\,,
 \label{lorenzo}
\end{equation} 
where $\Gamma > 0  ${.}
Inserting 
  the expression of $R(\omega)$ of 
Eq.~(\ref{eq:rs}) and  using the closure relation of the 
Hamiltonian eigenstates
\begin{equation} \label{compl1}
   \sum_n |\Psi_n \rangle \langle \Psi_n | = 1 \:\mbox{,}
\end{equation}
one finds that
\begin{equation}
  {L}(\omega_0,\Gamma)=\langle \Psi_0 | {{\mathcal O}}^{\dagger}\frac{1}{H-z^*}\frac{1}{H-z}{\mathcal O}|\Psi_0 \rangle \:\mbox{,}
\label{lorenzog}
\end{equation}
where $z=E_0+\omega_0+i\Gamma$.
Thus, in order to find the LIT, we need to solve the following Schr\"{o}dinger equation 
\begin{equation} \label{psi1}
  (H-z )|\widetilde{\Psi}\rangle =  {\mathcal O} | \Psi_0\rangle \:,
\end{equation}
for different values of $\omega_0$ and $\Gamma$.
The physical solution for the Lorentz states $|\widetilde{\Psi} \rangle$ of 
Eq.~(\ref{psi1}) has asymptotic boundary conditions like a bound-state. 
Moreover it is unique, due to the  fact that, because of the hermiticity 
of $H$, the homogeneous equation has only null solutions.
Once Eq.~(\ref{psi1}) is solved, one can get
 the transform as
\begin{equation} 
\label{elle}
  {L}(\omega_0,\Gamma )=\langle \widetilde{\Psi} |\widetilde{\Psi}
  \rangle \,,
\end{equation}
which can be
 evaluated
in a direct way, without requiring the knowledge of $R(\omega)$, nor the states $\Psi_n$
in the continuum.
The dynamical functions $R(\omega)$ is obtained instead by a numerical inversion of
the transform.  For details on the inversion procedure see, {\it e.g.}, Ref.~\cite{efros1999,andreasi2005}.

The great advantage of the LIT method is that it allows to avoid  the
complications of a continuum calculation, reducing the problem to the solution of a bound-state equation.
The interested reader can find more details in the review~\cite{Efl07}.
The LIT method has been benchmarked with alternative approaches, see, {\it e.g.}, the
two- and three-body systems in Refs.~\cite{efros1994,Sara} and has been applied with success on break-up observables
for nuclei with $A=4$ solving the Schr\"{o}dinger-like equation with NCSM~\cite{stetcu2007} and for nuclei with 
mass number $3\le A \le7$ solving the Schr\"{o}dinger-like equation with hyperspherical harmonics expansions.
Recent reviews of the method can be found in Refs.~\cite{Bacca2014,Efl07} and include many
examples with application to various electroweak reactions in light
nuclei. An alternative approach based on the idea of integral transform is provided by the Laplace transform, which is typically used in Green's function Monte Carlo methods, see, {\it e.g.}, Ref.~\cite{Lovato}.

\paragraph {Photodisintegration of six-body nuclei}
As an application of the Lorentz integral transform method used to tackle problems of relevance to the rare isotope physics,
below the case of the photodisintegration of six-body nuclei will be presented.

Nuclei with a number of nucleons 
between 4 and about 12 have traditionally constituted a bridge between the few- and the many-body systems.
For the mass  number $A=6$ the short-lived two neutron-halo nucleus $^6$He has received special attention both theoretically
and experimentally, in particular because it is the lightest of the halo nuclei.
The stable isotope among the $A=6$ isobars is instead the $^6$Li nucleus. In the study of dynamical properties such as response functions,   
an interesting 
question to pose is: does the interaction of a photon with these two isobaric 
analog nuclei lead to  different structures in the photodisintegration cross section? 

The photodisintegration cross section is defined as
\begin{equation} 
\sigma_{\gamma}(\omega)=4\pi^2 \alpha \omega R^{E1}(\omega)\,, 
\label{cs}
\end{equation}
where $R^{E1}$ is the dipole response function, {\it i.e.}, Eq.~(\ref{eq:rs}) where the excitation operator is the $E1$
dipole operator
\begin{equation}
{\mathcal O}=E1=\sum_{i=1}^A {\bf r}'_i\left(\frac{1+\tau_i^3}{2} \right)\,,
\label{dip}
\end{equation} 
 where $\tau_i^3$ is the third component of the isospin of the $i$-th nucleon, and consequently
$\frac{1+\tau_i^3}{2} $ is the isospin projector that selects only protons amongst nucleons.

\begin{figure}[t]
\begin{center}
\includegraphics[width=0.44\textwidth,clip=]{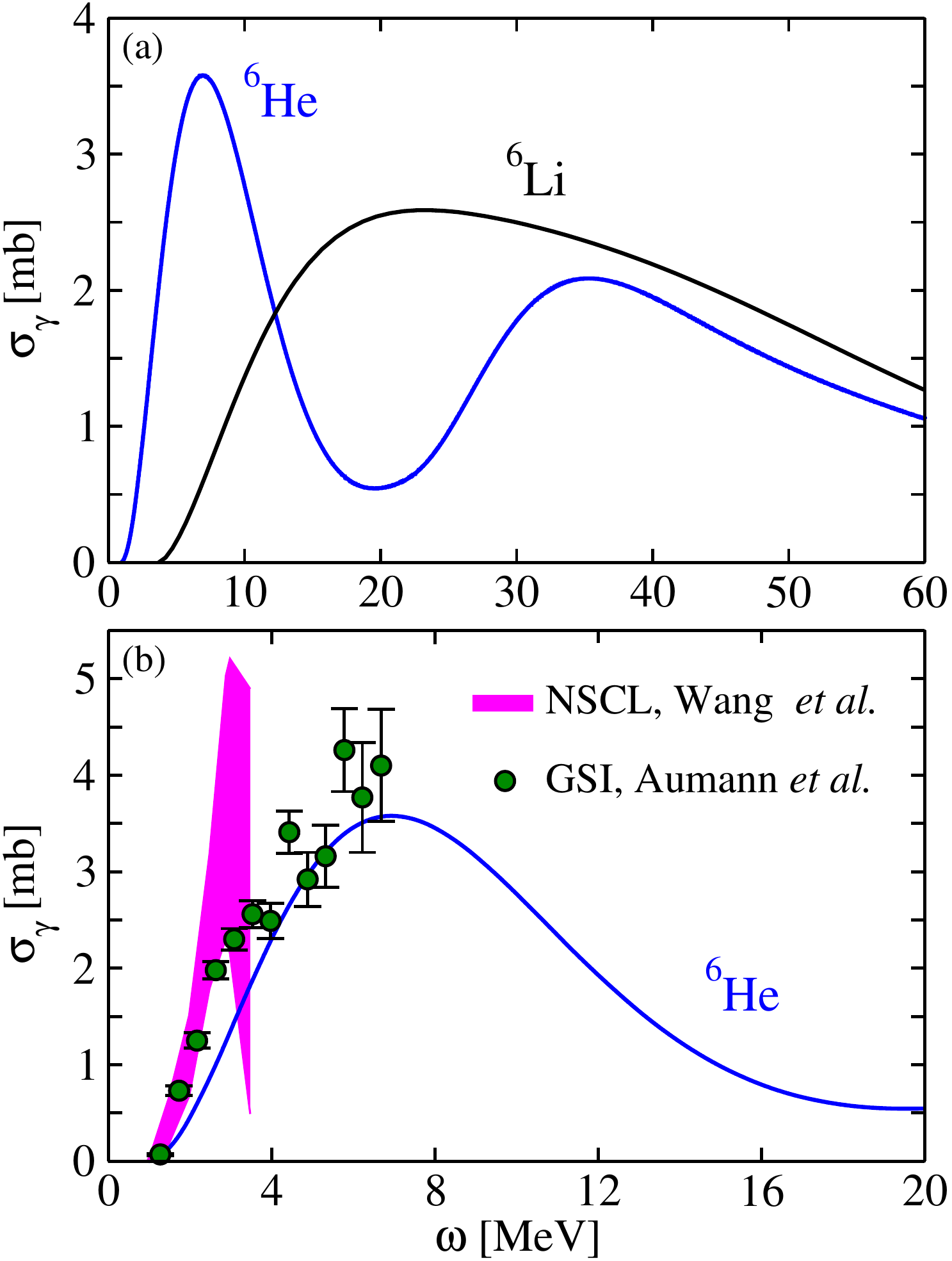}
\caption{(Panel (a): Theoretical photoabsorption cross section of $^6$He and $^6$Li from 
Ref.~\cite{bacca2002,BaB04} calculated with a semi-realistic potential. Panel (b): For the halo nucleus of $^6$He, data from Aumann {\it et al.}~\cite{aumann} and Wang {\it et al.}~\cite{Wang_diss} are shown in comparison to theory.}
\label{6photodis}
\end{center}
\end{figure}

The application of the LIT method used in conjunction with hyperspherical harmonics expansions
allows one to study the photodisintegration of the six-body nuclei and answer the above question.
In Refs.~\cite{bacca2002,BaB04} simple semi-realistic interactions 
were used to calculate $\sigma_{\gamma}$. Results are shown in Figure~\ref{6photodis}.
Such studies showed that
the halo structure of the rare $^6$He  isotope leads to considerable differences 
from the stable $^6$Li nucleus in photodisintegration. As shown in Figure~\ref{6photodis}(a), 
while a single resonant shape is observed for the cross section in $^6$Li,
in the case of $^6$He two
well separated peaks are seen.
The first peak corresponds to the break-up of the neutron halo, while the 
second peak corresponds to the break-up of the $\alpha$-particle, leading to a giant dipole resonance. Low-lying peaks observed in neutron-rich nuclei
are often also called pigmy dipole resonances. Different than for heavier nuclei, in case of $^6$He,  such low-lying 
fragmented strength is not small (pigmy), but is predicted by theory to be rather large.
The presence of two large peaks is robust and has been observed with three different semirealistic models of the nuclear force~\cite{bacca2002,BaB04}.

The $^6$He case was investigated
 with Coulomb excitation experiments performed at GSI by Aumann {\it et al.}~\cite{aumann} and at NSCL by Wang {\it et al.}~\cite{Wang_diss}. These data are shown in Figure~\ref{6photodis}(b). The two sets of data are consistent with each other in the energy range where they overlap and error bars are larger in the NSCL data.  One observes that theoretical results are describing the GSI data between 4 and 7 MeV rather well, even though some strength is missing at the very low energy. Given that the available experimental data only extend up to about 8 MeV of excitation energy, it would be really interesting to see if the presence of a low-lying separated peak as predicted by  theory can be confirmed experimentally.

\subsection{Lorentz integral transform with coupled--cluster theory}

The LIT method offers a great opportunity to study break-up observables avoiding the complication of continuum states. Until recently it was used only in conjunction with few-body techniques, such as hyperspherical harmonics or NCSM, thus restricting the range of possible studies to quite low mass number.
Thanks to the great advances made in coupled-cluster theory, it became evident that a coupled-cluster theory formulation of  the LIT method would pave the way for many new applications and studies of dynamical observables in the medium-mass regime. 
Recently, such potential was exploited and a new LIT formulation within coupled cluster theory (LIT-CC) was implemented~\cite{PRL2013,PRC2014}. Below, we briefly outline the strategy and present some exciting results.

In coupled-cluster theory, the response function is written as
\begin{equation}\begin{split}\label{respfunc}
R(\omega) = \sum_n &\langle 0_L|\bar{\Theta}^\dag| n_R\rangle\langle n_L|\bar{\Theta} |0_R\rangle\delta(E_n - E_0 - \omega),
\end{split}\end{equation}
where $\bar{\Theta}=e^{-T}{\mathcal O} e^{T}$  is the similarity transformed excitation operator and $\bar{\Theta}^{\dagger}$ its adjunct. The states
 $\langle 0_L|$, $|0_R\rangle$ are the left and right reference ground-states, with   $|0_R\rangle=| \Phi_0 \rangle$ and  $\langle 0_L|= \langle \Phi_0 | (1+\Lambda)$, where $\Lambda$ is a linear combination of particle-hole de-excitation operators~\cite{shavittbartlett2009}.
The states $\langle n_L|$, $|n_R\rangle$ are instead the left and right 
excited states, respectively.

The translationally invariant dipole operator of Eq.~(\ref{dip}) can be written as
\begin{eqnarray}
\label{dipole}
\nonumber
{\mathcal O}=E1&=&\sum_i^A \left({\bf r}_i -{\bf R}_{\rm CoM} \right) \left (\frac{1+\tau^3_i}{2} \right)\\
&=&\sum_i^A {\bf r}_i  \left (\frac{1+\tau^3_i}{2} \right) -\frac{Z}{A}{\bf r}_i\,,
\end{eqnarray}
which is a one-body operator and needs to be similarity transformed in coupled-cluster theory.

The Schr\"{o}dinger-like equation to solve becomes then
\begin{equation}\label{LITCC_eq}
(\overline{H} - z)|\widetilde{\Psi}_R\rangle = \overline{\Theta}|0_R\rangle\,.
\end{equation}
Solutions for $|\widetilde{\Psi}_R\rangle $
are found
by making the following {\it ansatz}
\begin{equation}
|\widetilde{\Psi}_R\rangle ={\cal R} | \Phi_0 \rangle \,,
\end{equation} 
where the operator  $\mathcal R$ is expanded in $p-h$ excitations  as
\begin{equation}
{\mathcal R}=r_0+\sum_{ia} r_{i}^a a_a^\dagger a_i+\frac{1}{4}\sum_{ijab} r_{ij}^{ab} a_a^\dagger a_b^\dagger a_j a_i\,.
\end{equation}
In other words, it is assumed that solutions for $\widetilde{\Psi}_R$  can be obtained as a linear combination of particle and hole excitation on top of the reference Slater determinant.

The formalism to solve Eq.~(\ref{LITCC_eq}) is analogous to an equation-of-motion~\cite{shavittbartlett2009} with a source in the right-hand-side and 
it has been implemented in Refs.~\cite{PRL2013,PRC2014,Orlandini14,Miorelli15}  with a truncation of both operators $T$ and ${\mathcal R}$ up to the $2p$--$2h$ excitation level, namely up to single and double (CCSD) excitations. 
 A benchmark with exact methods showed that, for $^4$He, the error introduced by the truncation scheme is of about $1-2\%$. Because coupled-cluster theory is size--extensive, similar errors are expected in heavier nuclei.
Inversions of the LIT are being performed as 
for the few-body calculations, using techniques described in Refs.~\cite{efros1999,andreasi2005}.

\begin{figure}[ht]
\centering
  \includegraphics[width=0.53\textwidth,clip=]{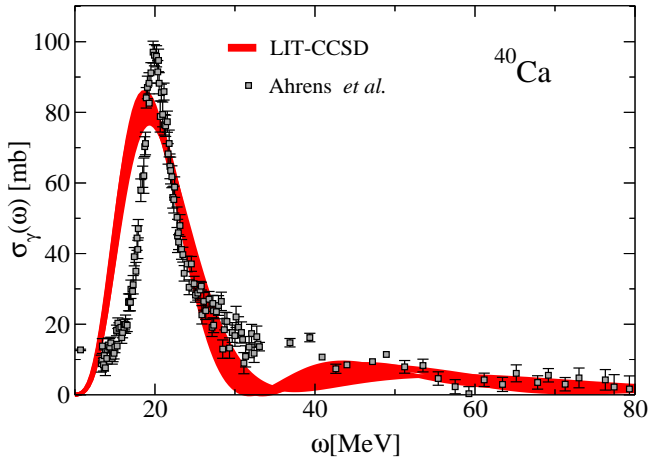}
  \caption{$^{40}$Ca photodisintegration
  cross section compared to  data from Ref.~\cite{Ahrens75}.  The curve is calculated with the N$^3$LO nucleon-nucleon force and is shifted to the  experimental threshold.}
  \label{fig_40Ca}

\end{figure}
Given that coupled-cluster theory scales mildly with mass number, it is possible to study medium-mass nuclei, such as $^{40}$Ca.
Results obtained with the LIT-CC method in the CCSD approximation, which we label LIT-CCSD, are shown in Figure~\ref{fig_40Ca}.
The LIT-CCSD approach opens up the possibility to investigate photodisintegration reactions 
using realistic nucleon-nucleon potentials, which reproduce two-nucleon scattering data. Results shown in Figure~\ref{fig_40Ca} are obtained
with a two-nucleon force from chiral effective field theory at next-to-next-to-next-to-next-to-leading (N$^3$LO) order~\cite{Entem03}.
The width of  the curve in Figure~\ref{fig_40Ca} is obtained by inverting LITs with different $\Gamma$ parameters in Eq.~(\ref{lorenzo}) and can be viewed as a lower estimate of the theoretical error bar.

As one can see, in Figure~\ref{fig_40Ca} the measured cross section by Ahrens {\it et al.}~\cite{Ahrens75}
shows  a very pronounced peak, referred to as the giant dipole resonance and located around  20 MeV of excitation energies.  This structure  is quite well reproduced by the LIT-CCSD theory. 

While first theoretical interpretations of such resonances,
 analogously observed in a variety of stable nuclei,
 were given in terms of collective models~\cite{GoT48,steinwedel1950}, most microscopic calculations available in the literature are based on
 mean-field approximations, see, {\it e.g.}, Ref.~\cite{Erler2011}. The LIT-CCSD method offers, for the first time, the opportunity to investigate such cross sections from first principles. 
Clearly, phenomenological mean-field approaches may give a better description of the data than present ab initio calculations. It is worth noticing, though, that the Hamiltonians presently used lack 3N forces. Work to include them in continuum calculations is underway and first results have been published in Ref.~\cite{naturephysics}.

\paragraph {Photodisintegration of neutron-rich nuclei} As an application of the LIT-CCSD method which is relevant to the physics of rare isotopes, we will present the photodisintegration of the neutron-rich $^{22}$O nucleus.

Photodisintegration reactions have been widely studied in the '70s with experiments on a variety of stable nuclei. More recently, at the rare isotope facilities, it has become possible to investigate analogous reactions for exotic nuclei, mostly via Coulomb excitation experiments. The comparison of stable and unstable nuclei can  provide key information about nuclear forces at the extremes of matter. Thus, it is very important to use  microscopically well founded  theories to address these observables.

\begin{figure}
\centering
  \includegraphics[width=0.54\textwidth,clip=]{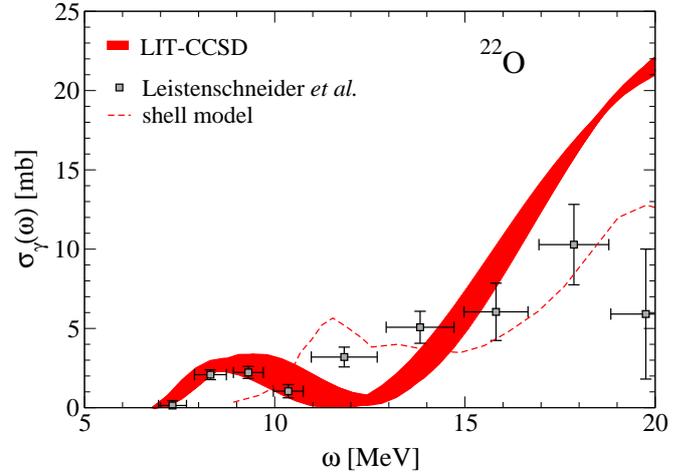}
  \caption{ $^{22}$O photodisintegration
  cross section compared with  data from Ref.~\cite{leistenschneider2001}.  The curve is calculated with the N$^3$LO nucleon-nucleon force and is shifted to the experimental threshold. The dashed line is a phenomenological shell model calculation.}
  \label{fig_22O}
\end{figure}

 The LIT-CCSD method enables one to study the photonuclear cross sections in neutron-rich nuclei with a closed shell nature. For example, neutron-rich oxygen isotopes have been  recently probed with Coulomb excitations experiments at GSI~\cite{leistenschneider2001}. The $^{22}$O  nucleus was studied in Ref.~\cite{PRC2014}  using the LIT-CCSD method with the same Hamiltonian as above 
and results are shown in Figure~\ref{fig_22O}.
It is very interesting to see that in the case of 
$^{22}$O  a small peak appears at low energy. The latter is often named pigmy dipole resonance and was experimentally observed in a variety of neutron-rich nuclei. With a first principle calculation and a two-body interaction which was tuned on two-nucleon data only, such substructure emerges naturally~\cite{PRC2014}. The theoretical cross section, when shifted to the experimental threshold, nicely agrees with data from Leistenschneider {\it et al.}~\cite{leistenschneider2001}. In the low-energy part, the agreement with experimental data is also superior to other calculations based on phenomenological shell model~\cite{leistenschneider2001} and also reported in Figure~\ref{fig_22O} by the dashed line. 

One observes that at higher energies the LIT-CCSD results is larger than data. This is expected because, while the experiment measured a semi-inclusive cross section where all neutrons were detected, the theoretical curve represents a total inclusive cross section, where  proton emission channels are also included.
Finally, it is worth noticing that no cluster structure has been imposed a priori in the calculation. The two-peak structure arises from a full microscopic calculation of twenty-two nucleons interacting with each other.
The pigmy dipole resonance observed in $^{22}$O resembles the $^6$He case discussed above, so it is expected to appear in other neutron-rich systems.
This is an example of new physical phenomena that arise in nuclear physics when
systems far from stability are studied. Other exotic nuclei are presently being investigated both theoretically and experimentally, such as $^8$He, $^{22}$C and $^{24}$O~\cite{tom}.

\section{Conclusions}
In these lecture notes we have reviewed microscopic models to describe the nucleus, starting from the historical non-interacting shell model approach and moving towards  some of the most sophisticated ab initio methods used in modern studies.
These notes do not purport to be a complete or exhaustive overview of all the progress done in the last 65 years in nuclear theory, but rather offer a short and easy introduction to this topical subject, which can be understood with simple quantum mechanical knowledge. The interested reader can find more information in some of the cited books, reviews or papers.

The writing up of these lecture notes was motivated by the summer school on Exotic Nuclei organized in Pisa, July 20th--24th, 2015~\cite{exotic2015}, which brought together over 80 undergraduate students from all over the world, gathered by the interest in the physics of rare isotopes. To them, and to any other interested non-expert reader, I would
like to reiterate the following take-home message:
Spurred by ideas and refinements of the non-interacting shell model picture, the modern  ab initio methods combined with $\chi$EFT approach offer the opportunity to link experimental observation with interactions at the fundamental level of quantum-chromo-dynamics. 

The rapid progress the theory nuclear structure and reaction has witnessed allows us to tackle new challenges and address increasingly complicated systems. Still many observations await a first principle explanation and novel challenges will be posed by the new data on exotic nuclei that will be collected at the rare isotope facilities.
\begin{figure}
\centering
  \includegraphics[width=0.8\linewidth]{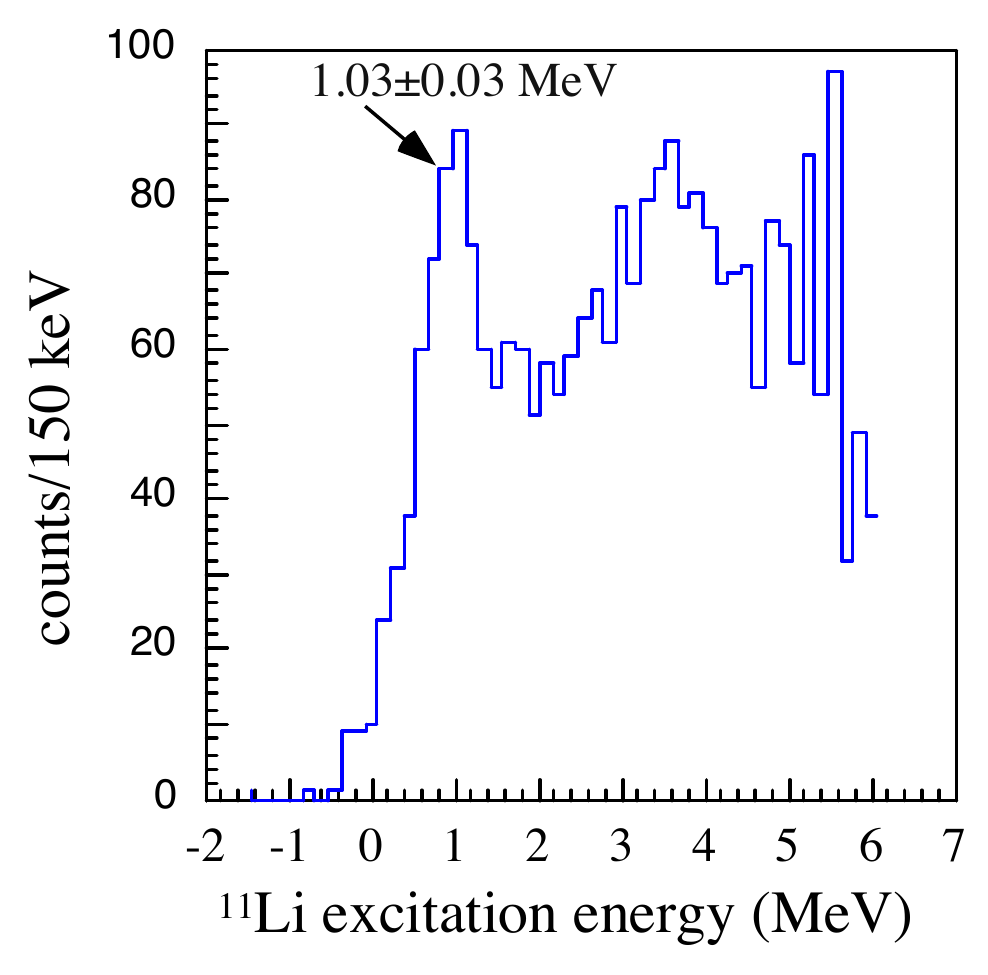}
\caption{Excitation energy spectrum from $^{11}$Li(d,d’). The isoscalar dipole resonance peak is seen at 1.03(03) MeV. Figure adapted from Ref~\cite{Ritu}.}
\label{ritu_fig}
\end{figure}
For example, recently, the world unique  IRIS facility at TRIUMF  made it possible to address the open question standing for two decades as to whether the soft dipole resonance, arising from an oscillation of the halo neutrons and the core, exists in the halo nucleus $^{11}$Li.   First evidence of dipole resonance with isoscalar character was observed from deuteron inelastic scattering, as shown in Figure~\ref{ritu_fig}. While modern shell model calculations indicate that the tensor force plays an important role, also first steps towards ab initio calculations are being taken. More exciting work is ahead of us to include three-nucleon forces and coupling to the continuum  and achieve a full microscopic understanding of these observations.

\vspace{1cm}
{\bf Acknowledgments}
I would like to thank the organizers of the summer school Exotic2015~\cite{exotic2015} for bringing together such a large number of young people interested in the physics of rare isotopes. I am grateful to M.~Miorelli, J.~D.~Holt and R.~Kanungo  for providing  some of their adapted figures. I would like to thank A.~Poves, J.~D.~Holt and S.~R.~Stroberg for useful discussions.
  This work was supported in parts by the Natural Sciences and Engineering Research Council (NSERC) and the National Research Council of Canada.

%

\end{document}